\documentclass[11pt, traditabstract]{aa}  
\graphicspath{ {./images/} }
\usepackage{graphicx}
\usepackage[varg]{txfonts}
\usepackage{longtable}
\usepackage{ulem}
\usepackage{hyperref}
\usepackage{threeparttable} % to use table notes
\usepackage{float}    % Recommended for better control over float placement
\usepackage{amsmath}
\usepackage{color}
\titlerunning{V CVn polarization} 
\authorrunning{Nikolov, Steenken, Ignace et al} 

\def\jobname{aa61677-26}
\begin{document} 

%%%%%%%%%%%%%%%%%%%%%%%%%%%%%%%%%%%

   \title{
Revealing the dusty shell of the semi-regular variable V CVn}

   \author{
          \href{https://orcid.org/0000-0003-0317-3866}{Yanko Nikolov}\inst{1,2}\fnmsep\thanks{\email{ynikolov@nao-rozhen.org}}
          \and
          \href{https://orcid.org/0000-0001-5744-4989}{Nicolaus Steenken}\inst{\ref{inst3}\fnmsep\thanks{ \email{nicolaus@steenken.info}}}
          \and
          \href{https://orcid.org/0000-0002-7204-5502}{Richard Ignace}\inst{4}\fnmsep\thanks{\email{ignace@etsu.edu}}
          \and
          \href{https://orcid.org/0000-0001-5415-9189}{Gail 
          Schaefer}\inst{\ref{inst5}}
           \and
          \href{https://orcid.org/0000-0002-3380-3307}{John D Monnier}\inst{\ref{inst6}}
          \and
          \href{https://orcid.org/0000-0001-9745-5834}{Cyprien Lanthermann}\inst{\ref{inst5}}
          \and
          \href{https://orcid.org/0000-0001-6017-8773}{Stefan Kraus}\inst{\ref{inst7}}
          \and
          \href{https://orcid.org/0000-0002-2208-6541}{Narsireddy Anugu}\inst{\ref{inst5}}
          \and
          \href{https://orcid.org/0009-0006-0225-4444}{Mayra Gutierrez}\inst{\ref{inst6}}
          \and
          \href{https://orcid.org/0000-0002-3003-3183}{Tyler Gardner}\inst{\ref{inst8}}
          \and
          \href{https://orcid.org/0000-0002-1788-9366}{Noura Ibrahim}\inst{\ref{inst6}}
          }
          
           \institute{
     Institute of Astronomy and National Astronomical Observatory, Bulgarian Academy of Sciences, 72 Tsarigradsko Chaussee Blvd., 1784 Sofia, Bulgaria \label{inst1}
     \and
     Nikola Vaptsarov Naval Academy, 73 Vasil Drumev str., BG-9002 Varna, Bulgaria \label{inst2}
     \and
    Sternwarte Freimann, Am Bluetenring 20, 80939 Munich, Germany (Amateur Astronomer)\label{inst3}
     \and
     Department of Physics \& Astronomy, East Tennessee State University, Johnson City, TN 37614, USA \label{inst4}
    \and
    The CHARA Array of Georgia State University, Mount Wilson Observatory, Mount Wilson, CA 91023, USA \label{inst5}
     \and
    Department of Astronomy, University of Michigan, 1085 S. University Ave, Ann Arbor, MI 48109, USA \label{inst6}
    \and
    Department of Physics and Astronomy, University of Exeter, Stocker Road, Exeter EX4 4QL, UK \label{inst7}
    \and
    Cooperative Institute for Research in Environmental Sciences, University of Colorado Boulder, Boulder, CO 80309, USA \label{inst8}
}
             
   \date{Received June 26, 2026 / Accepted October 21, 2026}
%%%%%%%%%%%%%%%%%%%%%%%%%

\abstract{
The semi-regular variable (SRV) star V~CVn exhibits a highly variable optical linear polarization, yet its position angle ($PA$) has remained remarkably stable at approximately $100^\circ \pm 20^\circ$ over several decades. We present a comprehensive multi-technique study of the system covering six pulsation cycles from 2020 to 2025. This campaign includes high-cadence photopolarimetry, spectropolarimetry, and the first interferometric observations of V~CVn using the CHARA Array (MIRC-X and MYSTIC) in 2024 and 2025.

Our photopolarimetric data reveal a significant increase in polarization, reaching peaks of up to 6\% during the 2023--2025 cycles, while the characteristic anticorrelation with brightness and the stability of the $PA$ persisted. The interferometric observations resolve the stellar photosphere but show no evidence of large-scale asymmetric structures, such as a bow shock or isolated scattering arcs, at the scales and sensitivities probed. 

Based on these results, we exclude previous models based on interstellar medium (ISM) interactions or static reflection clouds as the primary drivers of the observed polarization. Instead, we propose a circumstellar disk or flattened envelope—analogous to the $L_2$~Puppis system—as the most plausible explanation. We hypothesize that this stable geometry may be shaped or maintained by the star's high proper motion, where ram pressure from the ISM interaction favors a flattened distribution of circumstellar material. Within this stable framework, the polarization peaks are interpreted as the result of stochastic dust clumps, likely triggered by the star's long secondary period (LSP), which causes episodic increases in mass loss. This combination of an ISM-influenced disk geometry and transient, LSP-driven dust clumps naturally accounts for both the decadal stability of the position angle and the highly variable polarization amplitude of V~CVn.
}

   \keywords{
            techniques: interferometric; techniques: polarimetric;
            stars: AGB and post-AGB;
            stars: individual: V CVn (HD 115898);
            circumstellar matter;
            stars: mass-loss
            }
            
    \maketitle

%%%%%%%%%%%%%%%%%%%%%%%%%%%%%%%%%%%
\section{Introduction}

V Canum Venaticorum (V CVn) is a semi-regular variable red giant star characterized by a highly variable optical linear polarization. Remarkably, despite this high variability, the position angle of the polarization has remained mostly stable at $100^\circ \pm 20^\circ$ for decades, spanning approximately 100 pulsation cycles \citep{Neilson2014}. The primary pulsation period of the star is estimated to be between 189 and 194 days, although observations also suggest the presence of an extra-long secondary period of about 15 years \citep{Cadmus2024, Poliakova1989}.  Table \ref{tab1} shows a summary of its properties. 

The variable polarization of V~CVn was first identified in the 1960s by the pioneers of polarimetry \cite{Dyck1971} and \cite{1975ApJ...196..261S}. Subsequent measurement campaigns \citep{Davidson2014} have consistently confirmed that the star's brightness and its degree of polarization are anticorrelated. The combination of a fluctuating polarization amplitude—ranging from 0.3\% to $\sim 7$\%—and a fixed $PA$ implies a persistent non-spherical geometry. This stable orientation suggests a structural asymmetry that survives despite the star's significant optical variability of two magnitudes and the complex dynamics of its atmosphere. Such intrinsic linear polarization indicates an underlying asymmetry that, for V~CVn, could arise from atmospheric convection, wind bow shocks, or stable circumstellar structures.

Several models have been proposed to explain these unique polarimetric properties. \cite{Neilson2014, Neilson2023} suggested that the variable polarization is driven by a pulsation-driven spherically symmetric dusty wind that interacts with the interstellar medium (ISM). Given that Gaia data identify V~CVn as a high-velocity star moving at 95~km / s through the ISM, this interaction would create an asymmetric dusty shell or bow shock defined by the star’s velocity vector. In this scenario, dust forms during the brightness maximum, while subsequent expansion and interaction with the ISM produce the observed polarization maximum and the stable $PA$. \cite{Shrestha2021} simulated polarization arising from dust scattering within stellar wind bow shocks such as V CVn. They found that the polarized intensity is high at the bow head, and the largest fractional polarization arises from the lower part of the bow shock in the optical and JHK bands. 

An alternative model was proposed by \cite{Safonov2019}, motivated by differential speckle polarimetric observations. They introduced the concept of scattering arcs or reflection clouds located near the star. In this geometry, the variable polarization is explained by non-radial brightness variations of the star—such as dipolar pulsations or rotational modulation—which differently illuminate these static scattering structures over time.

However, the decadal stability of the position angle also points toward a more permanent, geometrically locked circumstellar environment. A compelling analogy can be found in the related SRV L$_2$~Puppis. Located at a distance of 56~pc, L$_2$~Puppis exhibits a similar pattern of polarization variability \citep{1986A&A...154....1M}. High-resolution VLT observations by \cite{Kervella2015} and \cite{Lykou2015} revealed that this variability is caused by a nearly edge-on circumstellar disk of dust and gas. The presence of such a disk provides a natural explanation for a fixed polarization axis that persists over long timescales. Given the similarities in their polarimetric signatures, a similar circumstellar disk or a flattened, stable dusty envelope could be a plausible cause for the observed phenomena in V~CVn, offering a robust alternative to the bow-shock and reflection-cloud hypotheses.

\begin{table}[ht]
\centering
\caption{Properties of V CVn (HD 115898)}
\label{tab1}
\begin{tabular}{@{}llll@{}}\hline
Parameter & Value / Range & Unit & Reference \\ \hline\hline
Distance & 499 +/-25 & pc & $^{a}$ \\
Tang.-velocity & 95 +/- 0.25 & km s$^{-1}$ & $^{a}$ \\
Pulsation period  & 191.89 & days & $^{b}$ \\
Spectral type & M6eIIIa -- M4e   &  & $^{b}$ \\
Brightness V & 8.6 -- 6.5 & mag & $^{b}$ \\
Brightness H & 2.7 -- 1.7  & " & $^{c}$ \\
Brightness K & 1.6 -- 1.1 & " &  $^{c}$\\
Photosphere radius& 430 +/- 40 & R $\odot$ &$^{d}$\\
Eff. temp. ($T_{\rm \text{eff}}$) & 3000 -- 3500  & K &$^{e}$ \\
Luminosity  & 1000 -- 3700 & L $\odot$ & $^{f}$\\
\hline\hline
\end{tabular}
   \tablefoot{
\tablefoottext{a}{\cite{2023A&A...674A...1G}}
\tablefoottext{b}{AAVSO and \cite{VSX2006}.}
\tablefoottext{c}{Converted from COBE DIRBE IR light curves \citep{2004yCat..21540673S}; see Appendix~\ref{Introduction}.}
\tablefoottext{d}{Estimates based on distance and $4\,\text{mas}$ diameter.}
\tablefoottext{e}{Estimates based on \cite{Sharma2016}.}
\tablefoottext{f}{Calculated via the bolometric distance modulus equation; see Appendix~\ref{Introduction}.}
}
\end{table}

The polarization variability of V CVn can thus be explained by different scenarios.  Understanding the morphology of the dust shell of V CVn will offer important insights into how evolved semi-regular variable stars lose mass and potentially explain their future evolution into planetary nebulae. 

To probe the asymmetric circumstellar envelope of V CVn, we obtained high-angular-resolution observations using the CHARA Array. To complement these spatial data and isolate the influence of variable intrinsic polarization, we present new spectropolarimetric observations alongside an estimate of the interstellar polarization baseline. In addition, we summarize a 6-year photopolarimetric monitoring campaign.

The paper is structured as follows: Section 2 details observations and data reduction. Section 3 outlines the data analysis, followed by Section 4, where we introduce a model for cycle-dependent polarization variations and compare its predictions against our data set. In Section 5, we discuss these results in light of current models from the literature before summarizing our main conclusions in Section 6.

%%%%%%%%%%%%%%%%%%%%%%%%%%%%%%%%
\section{Observations and data reduction} \label{obs}

\subsection{CHARA interferometric observations}
\label{CHARA}

V CVn was observed with the CHARA Array in 2024 and 2025. CHARA is a near-infrared interferometer that combines up to six telescopes with baseline lengths ranging from B = 34 to 331 m \citep{2005ApJ...628..453T}. In the wavelength range of the H band ($\lambda$ = 1.50 to 1.74 $\mu$m) and the K band (2.00 to 2.38 $\mu$m), it provides angular resolutions of approximately $\lambda$/2B 0.5 and 0.6 mas, respectively. We used the H-band MIRC-X \citep{Anugu2020} and K-band MYSTIC beam combiners \citep{Setterholm2022} in a cophasing observing mode.
The observations with CHARA in May 2024 were the first interferometric observations of V CVn and served as an initial determination of its diameter and signs of asymmetries. About a year later, another observation was carried out using five telescopes to obtain more detailed data of the asymmetric dusty shell. In total, V CVn was observed at three different epochs (Table \ref{CharaLog}) close to the brightness maximum. 

\begin{table}[!ht]
\caption{CHARA Array observing log of V CVn} \label{CharaLog}
    \centering
    \begin{tabular}{ l l l l l }
    \hline
        Date (UT) & Time (UT) & Combiner & Band & Telesc. \\ \hline\hline
        2024-05-08& 06:20, 7:34, 8:39 & MIRC-X & \ H& \ \ \ \  4$^{a}$ \\ 
        2024-05-08& 06:20, 7:34, 8:39 & MYSTIC & \ K& \ \ \ \  4$^{a}$ \\ 
        2024-05-09& 06:04 & MIRC-X & \ H&\ \ \ \   4$^{a}$ \\ 
        2024-05-09& 06:04 & MYSTIC & \ K& \ \ \ \ 4$^{a}$ \\ \hline
        2025-05-20& 04:27 & MIRC-X & \ H& \ \ \ \  5$^{b}$ \\
        2025-05-20& 04:27 & MYSTIC & \ K& \ \ \ \  5$^{b}$ \\\hline\hline
    \end{tabular}
    \raggedright
    \small
    \tablefoot{ $^{a}$ E1-E2-W1-W2, $^{b}$ S1-S2-E1-E2-W2}
\end{table}

We reduced the data using the standard MIRC-X / MYSTIC pipeline\footnote{\url{https://gitlab.chara.gsu.edu/lebouquj/mircx_pipeline}} (versions 1.4.0 and 1.5.0) with an integration time of 30~s. Data were calibrated using an IDL script written by J.~D.Monnier; the deep cleaning option was used to remove outliers, and data were averaged over 6-min intervals. The calibrator stars used are listed in table \ref{tab2}.

\subsection{Spectropolarimetric observations with FoReRo2}

Spectropolarimetric observations of V CVn and three stars in its direction were carried out using the 2-Channel Focal Reducer Rozhen (FoReRo2\footnote{Based on data collected with the Rozhen telescope, Bulgaria.}) \citep{2000KFNTS...3...13J}, mounted on the 2.0 m Ritchey–Chrétien–Coudé (RCC) telescope at the Bulgarian National Astronomical Observatory, Rozhen. A comprehensive description of FoReRo2 and its capabilities is presented in \cite{Nikolov2026}. 
The beam-swapping technique described by \citet{2009PASP..121..993B} was applied to minimize instrumental polarization. Two sets of polarization standard stars were used: unpolarized standards to correct for instrumental polarization and highly polarized standards to calibrate the position angle. Table \ref{tab.obsjournalspp} provides a summary of the spectropolarimetric observations.

\subsection{Photopolarimetry 2020 - 2025}
\label{Photopol}
 We continued our high-cadence polarization measurement campaign reported in \cite{Neilson2023} in the years 2023 to 2025. We used  dual-beam polarimeters with V-band filters mounted on custom Schmidt-Cassegrain telescopes in Munich/Germany (Observer 1) and Parton/Scotland (Observer 2). The polarimeter of Observer 1 had been upgraded 2023 as described in \cite{Steenken2025}. In total, we have observed six pulsation cycles since 2020. We have expanded the original catalog previously published in \citet{Neilson2023} by incorporating these new observations from 2023 to 2025. This update adds 153 new measurements to existing observations, resulting in a total of 380 polarimetric data points presented in Table~\ref{table:new_obs}.

\section{Data analysis} \label{Data analysis}
\subsection{CHARA interferometric observations}
\label{CHARA}
\begin{figure*}
    \centering
    \includegraphics[width=\textwidth]{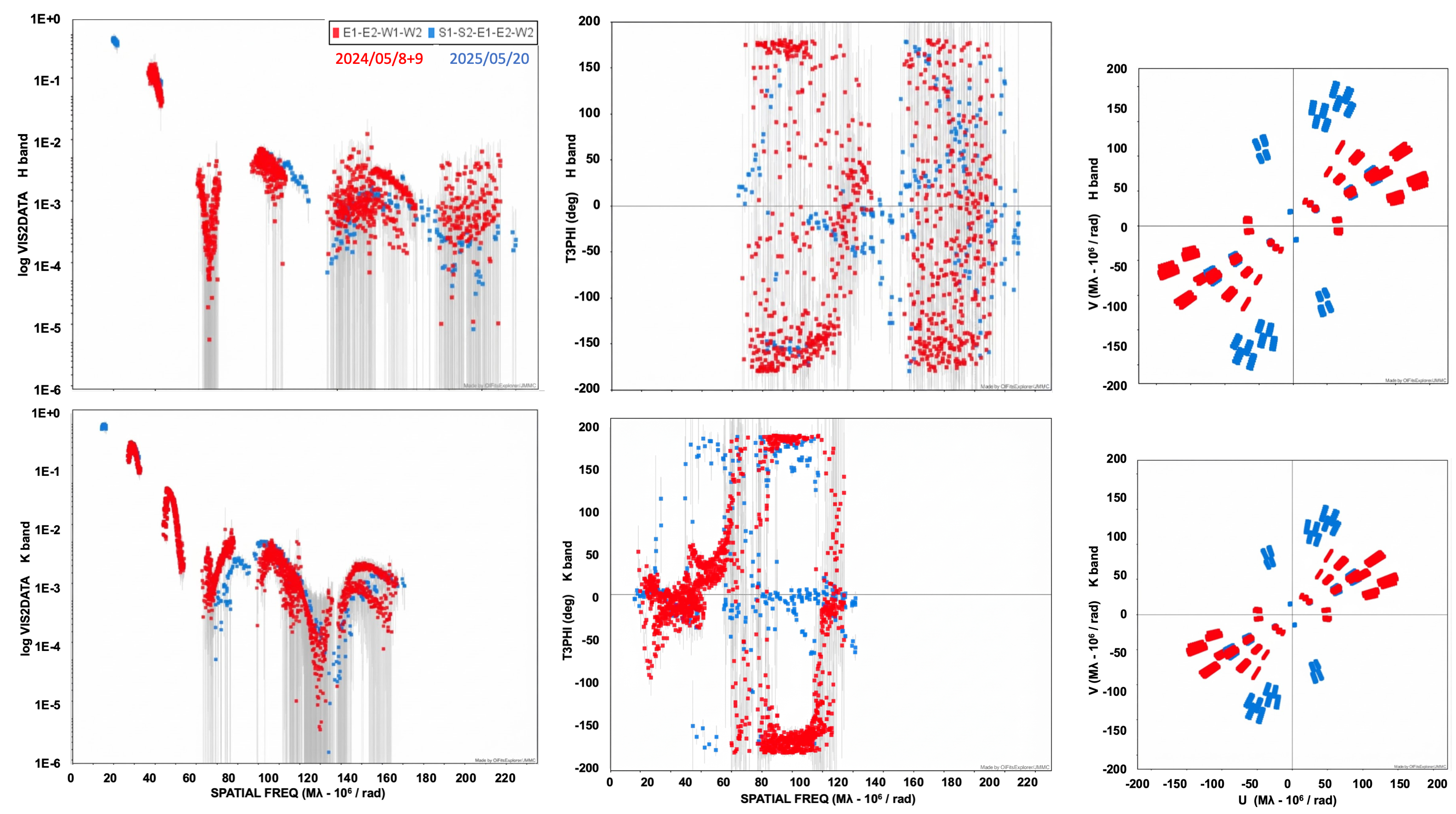} 
    \caption{Graphical representation of the reduced data sets from the three CHARA observations. The data from the two observations in 2024 are shown in red, the observation in 2025 in blue. The observations on May 8/9 2024  were made with four telescopes using the configuration E1-E2-W1-W2. The observation on May 20, 2025 was made with five telescopes in the configuration S1-S2-E1-E2-W2. The left column shows the squared visibilties in H band (upper panel) and K band (lower panel). The middle column exhibits the closure phases in H band upper panel and K band (lower panel). In the right column we show the coverage of the observations in the UV plane.} 
    \label{CHARA2025}
\end{figure*}

The visibilities and closure phases derived from the CHARA observations in 2024 and 2025 are shown in Fig. \ref{CHARA2025}.
The squared visibilities ($V^2$) (Fig.~\ref{CHARA2025}, left column) exhibit the characteristic profile of a resolved stellar disk, with the first minimum occurring at $\sim$ 66$~M\lambda$ in both bands. Using a geometrical model with a limb-darkened disk profile, we derive angular diameters of $3.8 \pm 0.15$ mas in the $H$-band and $3.95 \pm 0.06$ mas in the $K$-band continuum, with a limb-darkening power-law index of $0.25 \pm 0.10$. An analysis at the longest wavelengths of the $K$-band data revealed a decrease in $V^2$ within the CO line region. This localized drop in $V^2$ indicates that the star appears larger in the CO bands than in the continuum, consistent with the presence of an extended molecular layer. Beyond $2.31\,\mu\text{m}$, our CHARA/MYSTIC data exhibit a localized and significant drop in squared visibilities ($V^2$) and visibility amplitudes ($VISamp$) across the CO bandhead region ($2.31\text{--}2.39\,\mu\text{m}$). To evaluate the significance of this feature, the combined 2024 and 2025 CO-band data was fitted using a two-component geometric model consisting of a limb-darkened photosphere ($d_0$) surrounded by a concentric extended CO shell/ring. The best-fit continuum photosphere ($d_0 = 4.04 \pm 0.10\,\text{mas}$) is surrounded by an extended molecular ring of width $w_1 = 1.20 \pm 0.04\,\text{mas}$ contributing $\sim 20.5\%$ of the total flux in this spectral region.

The behavior of $V^2$  across the low spatial frequency regime provides a critical constraint on the presence of extended emission. In a multi-component system, significant off-axis structures or large-scale emission would lead to a measurable reduction in $V^2$  at the shortest baselines as that flux becomes over-resolved.

Beyond the radial profile, the closure phases (Fig.~\ref{CHARA2025}, middle column) provide  evidence for significant departures from point-symmetry. Although a symmetric source would produce closure phases restricted to $0^\circ$ or $\pm 180^\circ$, our observations show distinct deviations. In the H band, we measure significant non-zero closure phases between $\pm 100^\circ$ around $120$~M$\lambda$, with high-frequency data ($160$--$220$~M$\lambda$) scattered throughout the $\pm 180^\circ$ range. The K band data exhibit closure phases between +/- 50 degrees at low frequencies. At higher frequencies, there is a clear transition in $55$~M$\lambda$ reaching the $\pm 180^\circ$ range. These non-zero closure phases are indicators of asymmetries on and around the star, possibly arising from large-scale dust clumps or localized brightness spots on the photosphere. The differences between the 2024 and 2025 data points suggest intrinsic temporal variability in the circumstellar environment, although these changes must be interpreted with caution. Because the baseline position angles sampled nearly orthogonal quadrants between the two epochs a non-point-symmetric, static source would naturally yield different observables. The observed discrepancies likely reflect a combination of both true structural evolution and geometric projection effects, which cannot be definitively disentangled due to the incomplete UV-coverage.

Despite these observations, we have refrained from performing geometric model fitting or image reconstruction for these epochs. The primary limitation of the data is the incomplete Fourier plane (UV) coverage, leaving approximately one-third to one-half of the plane unsampled; specifically, the upper-left and lower-right quadrants (Fig.~\ref{CHARA2025}, right column). Incomplete UV sampling makes image reconstruction susceptible to artifacts and non-unique solutions that risk over-interpreting noise in the unsampled directions. The variability of the star between the two observation years also makes it difficult to combine and analyze the data from both observation years into a single data set.

Although the incomplete UV-coverage prevents image reconstruction, it does not affect the robustness of our asymmetry detection.  Because the observed closure phases deviate significantly from zero across several baseline triangles, centrosymmetry is mathematically ruled out. Unsampled spatial frequencies in the missing UV-quadrants cannot alter these already measured non-zero triple phases. Regarding what information is lost in the unsampled directions, the incomplete UV-coverage primarily restricts our ability to determine the exact morphological shape of the asymmetry. Furthermore, it prevents us from definitively disentangling the true structural evolution of the circumstellar environment between the 2024 and 2025 epochs from geometric projection effects caused by the different baseline orientations.

\subsection{Spectropolarimetric observations with FoReRo2}
\begin{figure}
   \centering
   \includegraphics[width=\linewidth]{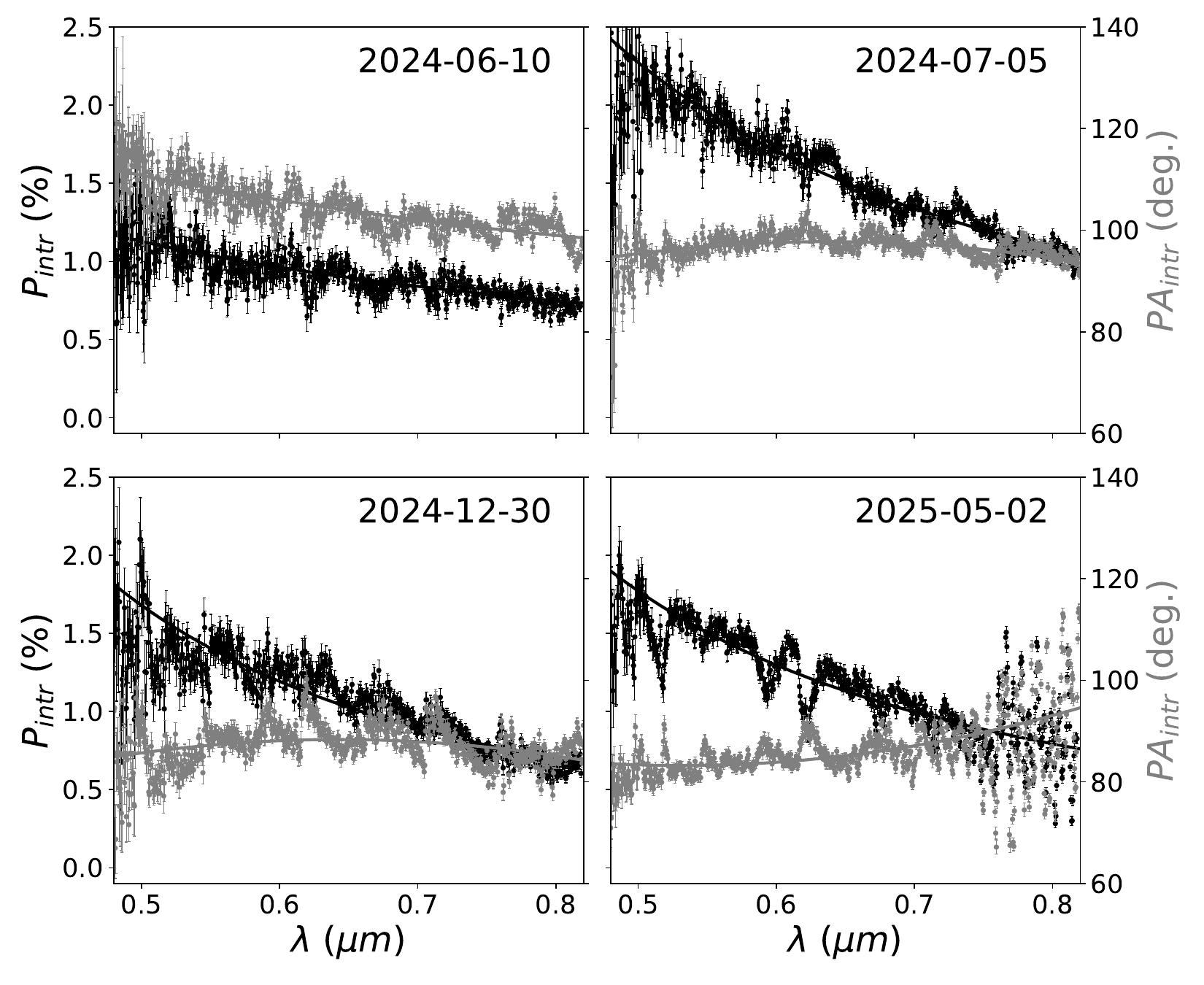}   
   \vspace*{0cm}
      \caption{Intrinsic degree of polarization (black) and $PA$ (grey) of the four measurements shown in \ref{tab.obsjournalspp}. The intrinsic polarization of V CVn increases towards shorter wavelengths. The $PA$ is not constant with the wavelength.}
         \label{VCVn.int}
   \end{figure}

The observed polarization is a vectorial sum of intrinsic polarization and interstellar polarization. A detailed description of the analysis of the interstellar polarization toward V CVn is presented in the appendix \ref{ISP_toward_VCVn}. The Stokes parameters of the intrinsic polarization $Q_{\rm \rm intr}(\lambda)$ and $U_{\rm intr}(\lambda)$ of~V CVn, were calculated using the following equations: 

\begin{eqnarray}
Q_{\rm intr}(\lambda)& =& Q_{\rm obs}(\lambda )-Q_{\rm isp}(\lambda ), \\
U_{\rm intr}(\lambda)& =& U_{\rm obs}(\lambda )-U_{\rm isp}(\lambda ),  
\end{eqnarray}
 
\noindent where $Q_{\rm obs}(\lambda)$ and $U_{\rm obs}(\lambda)$ are the observed Stokes parameters of V CVn. $Q_{\rm isp}(\lambda)$ and $U_{\rm isp}(\lambda)$ represent the Stokes parameters of interstellar polarization. Then we calculate the degree of intrinsic polarization ($P_{\rm intr}(\lambda)$) and the intrinsic position angle ($PA_{\rm intr}(\lambda)$).
The intrinsic degree of polarization (black) and the intrinsic $PA$ (grey) are presented in Fig. \ref{VCVn.int}. The $PA$ is not constant with the wavelength, but instead shows a smooth variation that can be modeled with a second-order polynomial function ($PA_{\rm intr} = a + b \times(\lambda/\lambda_0) + c \times (\lambda/\lambda_0)^2$), where $\lambda_0$ = 0.55 $\mu$m.
The second-order polynomial fitting coefficients are presented in Table~\ref{fit.coefficients}.

%-------------------------------------------------------------
%----------------------------------------------------------------

The intrinsic polarization of V CVn increases toward shorter wavelengths. To describe its wavelength dependence, we use a monotonic function in the form of a power law. The data are fitted using the following expression:
 
\begin{equation}
P_{\rm intr}(\lambda)=p_0\times\left(\frac{\lambda}{\lambda_0}\right)^{\beta}, 
\end{equation} 

\noindent where $p_0$ is a degree of polarization at $\lambda_0$, $\lambda_0$ = 0.55 $\mu$m, and $\beta$ is a power law index (the coefficients $p_0$ and $\beta$ are presented in Table~\ref{fit.coefficients}). The $\beta$ index reflects the nature of the scattering process: $i$) Mie scattering by dust grains, or $ii$) Thomson scattering by free electrons. The former is wavelength-dependent, while the latter is not. The power-law index $\beta$ decreases from $-0.87$ to $-1.88$ between 2024-06-10 and 2024-12-30.

\label{ISP}

\subsection{Photopolarimetry 2020 - 2025}
\label{Photopol}

\begin{table}[h]
\caption{Interstellar polarization (ISP) estimates for V-band ($5500$~\AA). Comparison between this work and \cite{Neilson2023}.}
\label{interstellarP}
\centering
\renewcommand{\arraystretch}{1.25}
\begin{tabular}{lcr}
\hline
Parameter & This work & \cite{Neilson2023} \\
\hline \hline
$P_{\rm isp}$ (\%)      & $0.29^{+0.13}_{-0.12}$ &  $0.34 \pm 0.05$ \\
$PA_{\rm isp}$ ($^\circ$) & $111.2^{+10.5}_{-14.1}$ &  $123.0 \pm 5.4$  \\
$Q_{\rm isp}$ (\%)      & $-0.22 \pm 0.07$ & $-0.14 \pm 0.10$ \\
$U_{\rm isp}$ (\%)      & $-0.20 \pm 0.10$ & $-0.31 \pm 0.09$ \\
\hline\hline
\end{tabular}
\end{table}

The observed images were calibrated and corrected for instrumental polarization as described \cite{Neilson2023} and used to calculate the calibrated Stokes parameters $Q_{\rm cal}$ and $U_{\rm cal}$. The intrinsic Stokes parameters $Q_{\rm intr}$ and $U_{\rm intr}$ were calculated from the observed Stoke parameters using the new interstellar polarization estimate which was derived in section 2.2 and shown in table \ref{interstellarP} analogous to equations (1) and (2).

In the next step, we calculate the intrinsic polarization $P_{\rm intr}$ and the intrinsic polarization angle $PA_{\rm intr}$ according to:
%-------------------------------------------------------------
   \begin{figure}
   \centering
   \includegraphics[width=\linewidth]{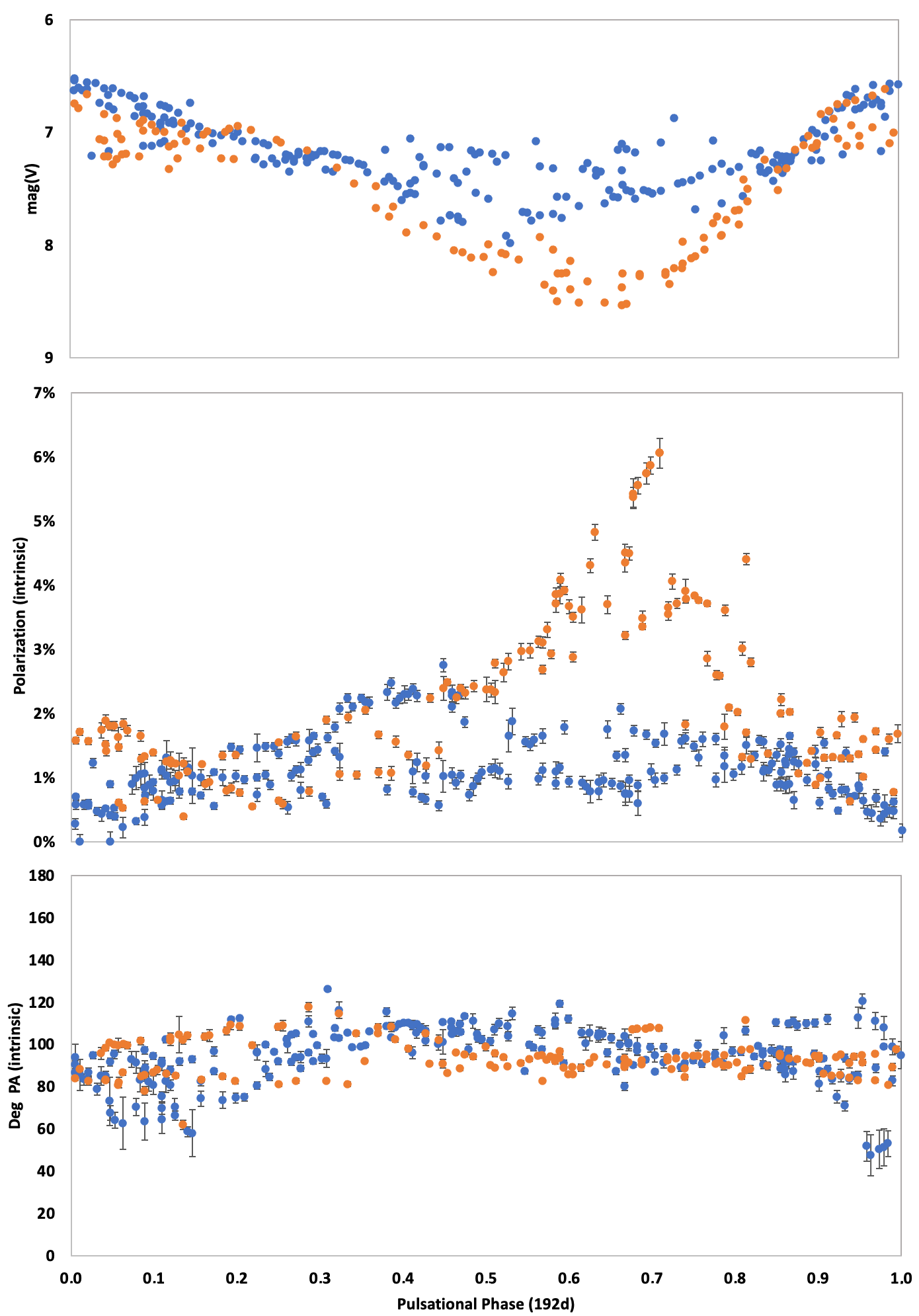}
   \vspace*{0cm}
      \caption{Phase-folded variations of photometric brightness (top), the intrinsic linear polarization $P_{\rm intr}$ (middle) and position angle $PA_{\rm intr}$ (lower panel). The data cover six pulsation cycles from 2020 to 2025, folded with a period of P=192 d. Phase 1.0 corresponds to the brightness maximum at JD 2459060 (2020-07-29). The cycles are color-coded to distinguish between two activity states: periods from 2020 to 2022, characterized by a maximum polarization below 3\%, are shown in blue, while the 2023–2025 cycles, exhibiting higher peak polarizations between 4\% and 6\%, are shown in orange.}
         \label{VCVn.IntrPhase}
   \end{figure}
%----------------------------------------------------------------
\begin{eqnarray}
P_{\rm intr} & = & \sqrt{Q_{\rm intr}^2+U_{\rm intr}^2}, \\
PA_{\rm intr} & = & \frac{1}{2}\,\tan^{-1} \left(\frac{U_{\rm intr}}{Q_{\rm intr}}\right).
\end{eqnarray}
The averaged error margins of intrinsic polarization  for all measurements 2020 - 2025 were +/-0.15\%, the averaged error margins of the PA +/-$5.2^\circ$ (2020 - 2022) and due to upgrades of polarimeters +/-$2.5^\circ$ from 2023 - 2025. The calculation of the errors is described in Appendix C. 

Calibration stars were observed to determine the stability and reliability of the polarization position angle. 

The V-band photometry of V~CVn was obtained directly before or after the polarimetric measurements with nearby comparison stars. Ob1 selected the nearby star BD +46 1863, with $mag(V) = 10.28$, as a comparison. Ob2 used an ensemble of the four AAVSO comparison stars \footnote[3]{Chart X28069ADU, AAVSO Variable Star Database}.

The polarization curve during the pulsation phase (Fig. \ref{VCVn.IntrPhase}) shows significant differences between the six observed pulsation cycles. Although the cycles in the first observation period showed maximum polarization strengths of 3\% around the brightness minimum, the polarization in the second observation period (2023 to 2025) reached up to 6\% at phase 0.7. Away from the brightness maxima, the polarization angle (PA) moved within a relatively narrow band of $100^\circ$ ± 20. The variability of both polarization and PA during the first period (2020–2023) was predominantly caused by the pulsation period observed in 2021, during which intrinsic polarization was significantly higher and PA dropped below $60^\circ$ during the polarization minima.   Figure \ref{Historic} shows the long-term light and polarization curves since 1966. Since the beginning of polarization measurements, polarization above 5\% has only been measured in three phases in the last 60 years. That is, in 1966 by \cite{Serkowski1966} and \cite{Serkowski2001} and around 1995 by \cite{Davidson2014}. The brightness minima and maxima in Fig. \ref{Historic} show multi-year secondary periods. This suggests that the star has either an extra-long second pulsation period in which significantly more dust is formed or that dust clouds orbit and darken the star. The observation of particularly high polarization values during a pronounced minimum brightness suggests that these phenomena are linked. Polarization maxima were observed at the same time as brightness minima, but also some days or weeks later or earlier, as shown in Fig. \ref{IntrinsicPol}. 

%%%%%%%%%%%%%%%%%%%%%%%%%%%%%%%%%%%
\section{Considerations for a polarization model} \label{Model}
%(Rico)

In order to capture the basic observed properties of the polarimetric variability in the visible optical band, we introduce a phenomenological model in hopes of clarifying the attributes that better and future physical modeling must take into account. The main characteristics of the observations can be summarized as follows:

\begin{itemize}
\item The scattering region is relatively compact on the basis of CHARA data, on the order of several stellar radii.
\item The polarization position angle is generally long-term steady with only modest variations during the pulsation cycle
\item  The model needs to reproduce that the PA stays within the observational constraints
\item There are slight variations in the PA and in relation to the polarimetric light curve as not being the same from one cycle to the next.
\item The interstellar polarization is largely negligible.
\item The polarization amplitude can vary significantly, from around 1\% to in excess of 6\%. 
\item The polarization changes with wavelength as well (``chromatic''). However, the wavelength dependence is not a primary consideration for modeling here.
\end{itemize}

Taking these points into account, we consider a simplistic model with the following properties.

\begin{itemize}
\item Optically thin dipole scattering is assumed, which allows us to employ approaches developed by \cite{1977A&A....57..141B} and \cite{1987ApJ...317..290C} for the polarization of scattered light.

\item A long-term axisymmetric envelope configuration. The number density of scatterers, $n(r,\theta)$, is taken to be of the form:

\begin{equation}
n(r,\theta) = \left( \frac{R_\ast}{r}\right)^2\,\left( a + b \sin^m\theta\right),
\end{equation}

\noindent where $r$ is radius in the envelope from the star center, $a$ and $b$ characterize the relative shape of the axisymmetric envelope, and $m$ is a power-law exponent for how density changes with co-latitude. Note that if $b=0$ or $m=0$, the envelope is spherical.

\item To account for variability, wind clumps are invoked. The presence of inhomogeneities (``clumps'') in stellar winds is well known, from hot massive stars \citep[e.g.,][]{2008A&ARv..16..209P} to cool evolved ones \citep[e.g.,][]{2023A&A...669A.155F}. Our model allows for the random injection of spherical scattering clumps that compete with the long-term stable component of the distorted circumstellar envelope.

\end{itemize}

For the initial exploratory and phenomenological model, we employ the approach of \cite{1977A&A....57..141B}: dipole scattering (Rayleigh-like) in an axisymmetric circumstellar envelope as illuminated by a point star. The point-star assumption is known to overpredict polarization \citep{1987ApJ...317..290C}; however, our goal here is to match observed trends, not quantitatively fit specific datasets.

The polarization of \cite{1977A&A....57..141B} is of the form

\begin{equation}
p_0 = \tau_0(\lambda) \, (1-3\gamma)\,\sin^2 i,
    \label{eq:p0}
\end{equation}

\noindent where $p_0$ is the polarization from the spatially unresolved envelope with the symmetry axis viewed at an inclination angle $i$. The parameter $\gamma$ is known as the ``shape factor'' and can be interpreted as a moment calculation over the scattering envelope. Its value ranges from $-1$ for a narrow bipolar jet to $0$ for a spherical envelope to $+1$ for a planar equatorial disk. Finally, $\tau_0$ is an angle-averaged envelope optical depth, which can be a function of wavelength \citep[.g., Rayleigh scattering scales as $\lambda^{-4}$ in the optical and infrared][]{2025Ap&SS.370..137I}.

The shape factor is calculated through the following

\begin{equation}
\gamma = \frac{\tilde{N}}{\bar{N}},
\end{equation}

\noindent where following \cite{2022ApJ...933....5I},

\begin{equation}
\bar{N} = \int\,n(r,\mu)\,dr\,d\mu,
\end{equation}

\noindent and

\begin{equation}
\tilde{N} = n(r,\mu)\,dr\,\mu^2\,d\mu,
\end{equation}

\noindent where $\mu = \cos \theta$. Note that a variable polarization at fixed PA can be achieved by allowing the parameters $a$, $b$, and/or $m$ to vary, possibly as functions of the pulsational phase. For axisymmetry it is possible that evolution of these parameters could alter the shape function $\gamma$ leading to a change of sign in equation~(\ref{eq:p0}). In this model such a sign change corresponds to a rotation of $90^\circ$ in the polarization PA.

%Note that a PA rotation of $90^\circ$ for certain changes in parameters, but only by that amount.

To allow for more mild variations in PA, as well as polarization, we consider random and evolving scattering clumps in the envelope. The approach is similar to that of \cite{2023MNRAS.519.3271I} in the context of a massive star wind. Clumps are injected into the wind along radial trajectories from the star. The direction $\theta$ and $\phi$ are random. There is an injection rate or frequency, $\dot{N}$, that is constant in time average but randomly sampled on characteristic time interval, $\delta t$. 

For the computational process, let $\{s\}$ be a set of random numbers, each value between 0 and 1. In the following each occurrence of $s$ is a different value. Random directions for clumps mean random in solid angle, with

\begin{equation}
\mu_{\rm j} = -1 + 2\, s,
\end{equation}

\noindent and 

\begin{equation}
\phi_{\rm j} = 2\pi\,s.
\end{equation}

\noindent where $j$ is the j$^{th}$ clump. Random but uniform sampling in time is achieved by introducing the clump injection time

\begin{equation}
t_{\rm j} = T \, s,
\end{equation}

\noindent where $T$ is the time duration of a simulation. Then for a number of clumps ${\cal N}_{\rm cl}$, the average time interval between clumps is $\delta t = T/{\cal N}_{\rm cl}$, and the average injection rate of clumps is $\dot{\cal N}_{\rm cl} = 1/\delta t$.

Each clump is treated as optically thin to scattering. As long the clump has relatively small extent in solid angle, its exact shape is immaterial \citep[e.g.,][]{1995A&A...295..725B}. A single isolated clump is thus approximately an axisymmetric density structure as defined by its vector trajectory through location $(\mu_{\rm j}, \phi_{\rm j})$. Consequently, the construction of \cite{1977A&A....57..141B} applies, but with an inclination $i_{\rm j}$ relative to the observer axis given by

\begin{equation}
\cos i_{\rm j} = \sin\theta_{\rm j}\,\cos \phi_{\rm j}.
\end{equation}

\noindent The position angle of a clump about the observer axis, using spherical trigonometry relations, is given by

\begin{equation}
\tan \psi_{\rm j} = \frac{\cos\theta_{\rm j}}{\sin\theta_{\rm j}\,\sin \phi_{\rm j}},
\end{equation}

\noindent written explicitly as numerator over denominator so that quadrant ambiguity may be resolved, since $\psi \in [0,2\pi]$.

For simplicity, clumps are assumed to move at constant speed, such that their radius as a function of time, $t$, is given by

\begin{equation}
r_{\rm j}(t) = R_\ast + \frac{t-t_{\rm j}}{v_{\rm cl}},
\end{equation}

\noindent where $t_{\rm j}= s_{\rm j} \, T$, and $v_{\rm cl}$ is the constant radial speed of the clump. We take all clumps to move at the same speed, an assumption which could be relaxed.

Treating the clumps as optically thin, the polarization contributed by an individual clump becomes

\begin{equation}
p_{\rm j}(t) = \tau_{\rm cl,j}(r_{\rm j})\,\sin^2 i_{\rm j}.
\end{equation}

Each clump contributes a polarization to the total that depends on its Stokes parameters, with

\begin{flalign}
q_{\rm j} (t) &= p_{\rm j}\,\cos (2\psi_j), & \\
u_{\rm j} (t) &= p_{\rm j}\,\sin (2\psi_j) &
\end{flalign}

\noindent which applies for those clumps with $t_{\rm j} \ge t$.
Under that same condition, the total polarization at a given time, including the envelope, plus the ensemble of clump up to that time, is given by the relationships:

\begin{align}
& q(t) = q_0(t) + \sum_{\rm j} q_{\rm j}(t), \\
& u(t) = u_0(t) + \sum_{\rm j} u_{\rm j}(t), \\
& p(t) = \sqrt{q(t)^2+u(t)^2}, \\
& \tan (2 \psi(t)) = \frac{u(t)}{q(t)}.
\end{align}

The simple superposition of all these contributions requires that every sightline through the envelope be optically thin. This would need to be verified with realistic envelope parameters. However, the value of this construction is to permit basic lightcurves in polarization with time to evaluate the extent to which physically plausible wind inhomogeneities might add ``noise'' to the otherwise stable variable polarization of the envelope. Optical depth effects will not alter the basic conclusions of combining a random component with a long-term stable one.

\begin{figure}[t]
   \centering
   \includegraphics[width=\columnwidth]{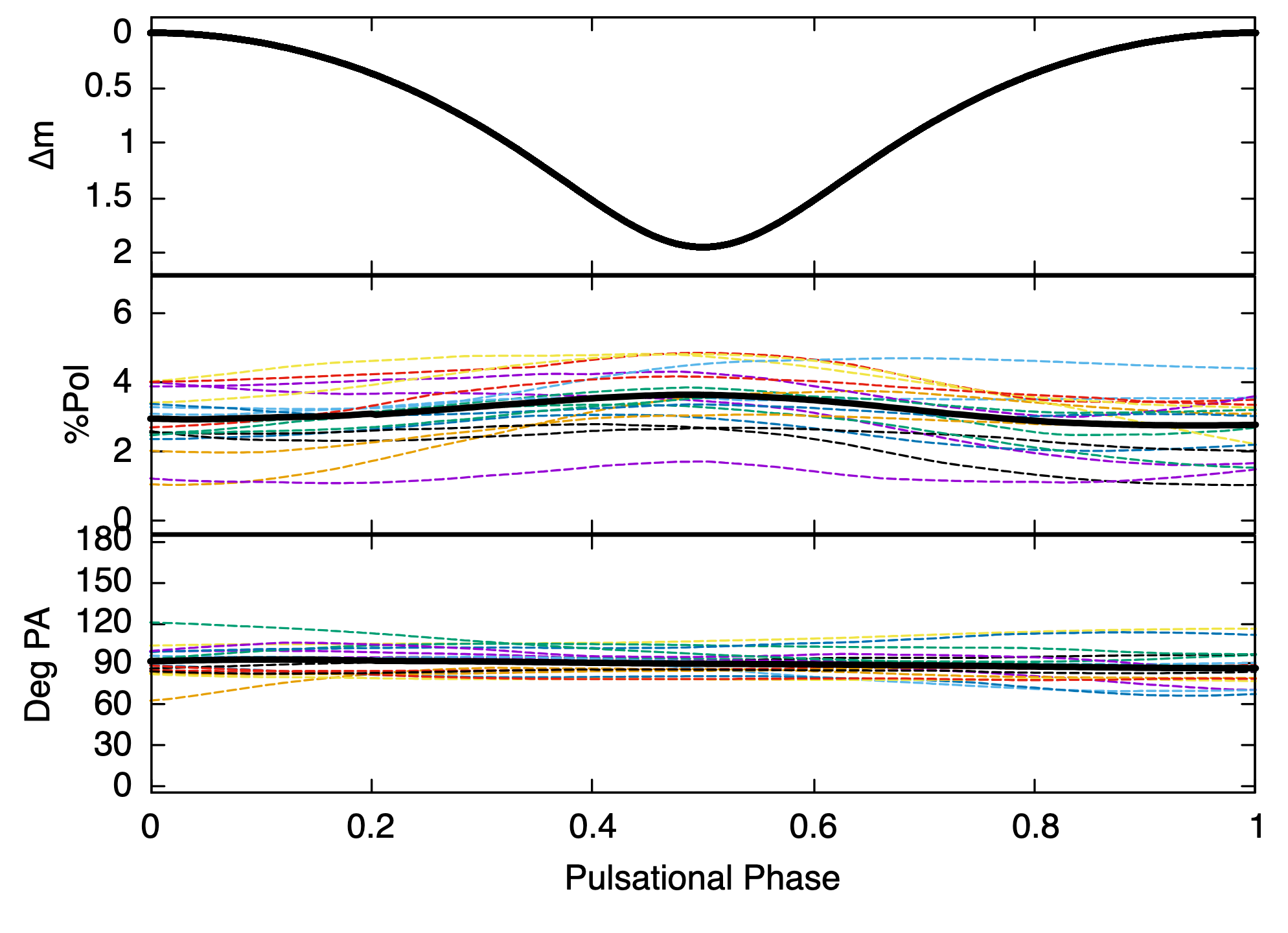}
     \caption{Results of 18 separate model calculations of optically thin dipole scattering for magnitude (top), polarization (middle) and PA variations with pulsation phase. 
     The black curves for polarization and PA are the average values. The dotted lines are for the 18 separate model runs, each with 200 clumps randomly distributed over a radial extent of $\approx 56 R_\ast$. The presence of clumps produces significant changes in the polarization, with an amplitude of around half the stable signal. In some pulsational cycles the polarization maximum occurs before or after brightness minimum like in the observed cycles (\ref{IntrinsicPol}). The PA shows variations of $\sim 20^\circ$ from the stable value. } 
    \label{fig:model}
   \end{figure}

Here, our intention is to select a set of model parameters that appears to reproduce the observed characteristic variations. Figure~\ref{fig:model} shows a selection of model examples. The luminosity of the star governs the relative flux, for which we use:

\begin{equation}
L \propto \frac{1 + 5\,\sin^2 (\pi\,t/P_{\rm puls})}{6},
\end{equation}

\noindent where $P_{\rm puls}$ is the pulsation period of the star. We have chosen the star to vary between a normalized maximum of unity, with a minimum of $1/6$, which is approximately the case for V CVn. The relative flux is the top panel of Figure~\ref{fig:model}.

Below that in the middle is the total polarization, with the bottom showing the PA. We have chosen $b=3$ and $m=2$ as fixed. The envelope parameter $a \propto L$ in that it varies from 1/6 to 1. With $m=2$, the envelope shape factor becomes $\gamma = (a+0.4b)/(3a+2b)$, which is a function of time. Then we choose $p_0 = 10\,(1-3\gamma)$. This assumes a fixed inclination of $i_0 = 90^\circ$ to maximize polarization. It also requires a large multiplicative coefficient to obtain the high levels of observed polarization. The value of $\gamma$ ranges between about 0.21 and 0.24, and it is assumed that each clump has the same optical scale of 0.4. The speed of the clumps is such that they move $R_\ast$ to $3R_\ast$ in half a pulsational phase. This is around $860 R_{\odot}$ in about 95 days, corresponding to a speed of $\sim 75$~km/s. Following \cite{2023MNRAS.519.3271I} for the basic form, the contribution of a clump to the polarization is taken to be

\begin{figure}[t]
   \centering
   \includegraphics[width=\columnwidth]{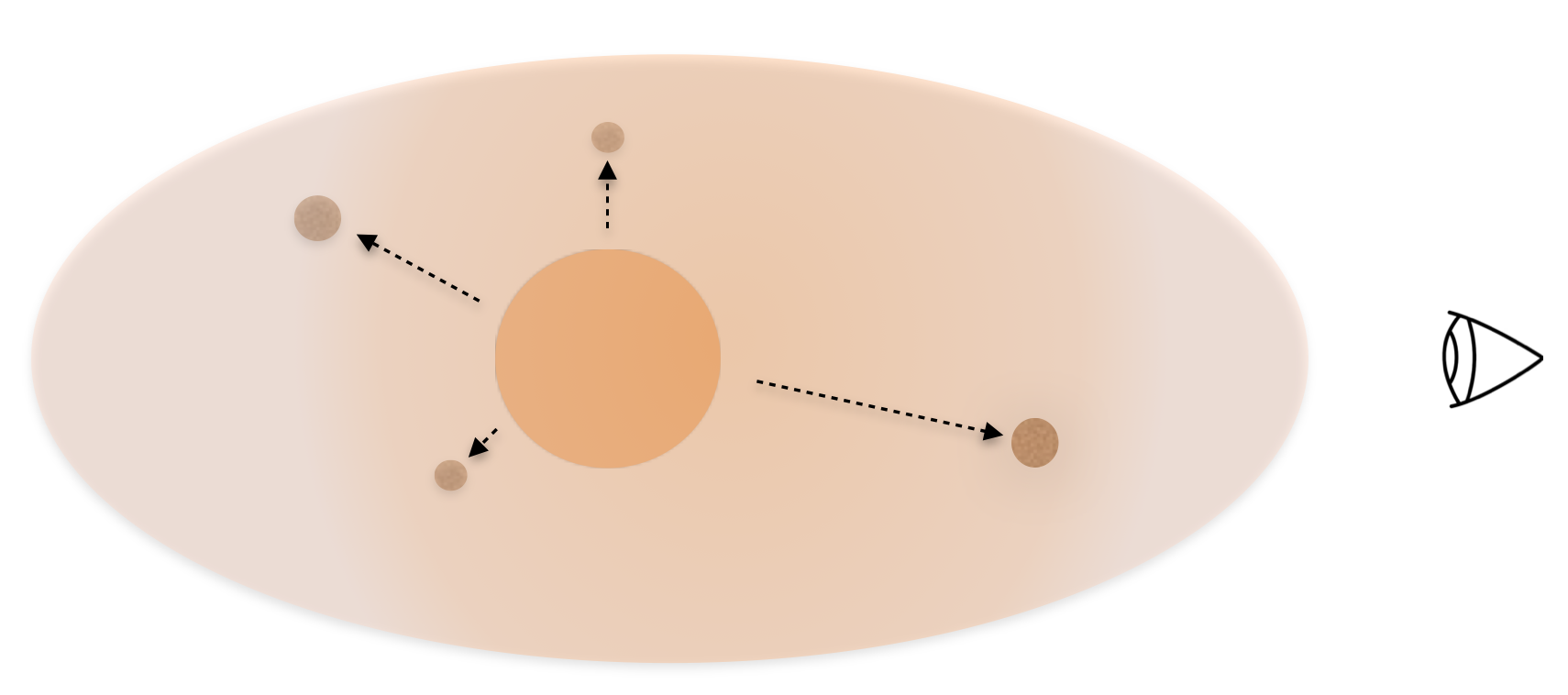}
     \caption{Schematic illustration of the model geometry. A central star (radius $R_\ast$) is embedded in an axisymmetric dusty envelope with a co-latitude-dependent density profile. Four optically thin small dust clumps are shown moving radially outward from the photosphere to $3R_\ast$ at a constant velocity. The eye symbol on the right indicates the observer's line-of-sight.} 
    \label{fig:scetch}
   \end{figure}
   
\begin{equation}
p_{\rm cl} \propto \left(\frac{r}{R_\ast}-1\right) \,e^{-\left(r/3.5R_\ast\right)^2}.
\end{equation}

\noindent This form ensures that the polarization of a clump is zero right at the star's photosphere, peaks around $3R_\ast$, then declines exponentially thereafter. Studies  that dust grain growth at around this radius motivates the choice of $3R_\ast$ \citep{Ohnaka2016, Aronson2017, 2023A&A...669A.155F}.

Figure~\ref{fig:scetch} shows a schematic illustration geometry of the model and figure~\ref{fig:model} shows the
results of 18 independent model runs. Each model involves the injection of 200 clumps ``seeding'' the wind by allowing them to move out radially over a span of 7 pulsation periods. Thus, the average time of clump injection is approximately weekly on average, with actual clump injection randomized as uniformly sampled over the 7 pulsational periods. Results from the simulation are shown for the pulsational phase over the final pulsational period. 

The overall level of polarimetric variation along with variations in PA from the long-term stable value are consistent with the observations. However, the model has several major deficiencies. Given the model parameters involved, the ensemble of clumps may not be optically thin, unless their distribution is quite spatially sparse to avoid overlapping clumps projected onto the plane of the sky. Although the non-zero closure phases indicate a central brightness distribution that is asymmetric, we have no physical basis to understand that shape. Thus, our model is quite phenomenological. At least it gives a sense for what scale of parameters will be necessary in terms of envelope distortion and a random contribution in order to reproduce the observations.

\section{Discussion}
\label{discus}
The decades-long stability of the position angle ($PA \approx 100^\circ$) in V~CVn, which has persisted through strong changes in polarization amplitude and optical brightness, requires a long-term asymmetric structure in the circumstellar environment. Our multi-technique study, especially the high-angular-resolution data from the CHARA Array, allows for a critical re-evaluation of previously proposed hypotheses.

The bow-shock model proposed by \citet{Neilson2014, Neilson2023} assumes that the interaction between a spherically symmetric stellar wind and the interstellar medium (ISM) creates the necessary asymmetry. Unlike the many AGB stars for which bow shocks have been successfully identified in the infrared, V CVn remains an exception with no such feature detected to date. V CVn falls outside the scope of the infrared survey by \cite{2012A&A...537A..35C}; as such, it is not listed among either the detected bow shocks or the documented non-detections. Based on the star’s distance and mass-loss parameters, a potential bow 
shock would be located at a distance of more than $1000\,\text{AU}$. This corresponds to an angular separation of $\gtrsim 2^{\prime\prime}$ from the central star. Although such a structure would be encompassed within the aperture of our dual-beam polarimetric observations ($> 2^{\prime\prime}$), it is far beyond the scales probed by our 
interferometric observations. Crucially, light scattering from such a remote, extended structure would be far too faint and could not follow the 192-day pulsation cycle in the synchronized manner observed. Thus, while a bow shock might exist within our photometric field, it cannot be the driver of the high and rapidly varying intrinsic polarization.

Similarly, the hypothesis of scattering arcs or ``bipolar clouds'' \citep{Safonov2019} appears to be difficult to reconcile with our results. This model posits static reflection nebulae located at an angular separation of $\sim 35\,\text{mas}$ from the star. This separation falls well within the field of view of the CHARA Array. However, the datasets reveal no clear signature of discrete, high-contrast scattering structures. Although our modeling framework is specifically designed to tightly constrain the inner circumstellar environment within a few stellar radii where the bulk of the dust forms, we can evaluate the geometric viability of these more distant arcs. In the refined geometric configuration of \citet{Safonov2019}, these structures are represented by discrete arcs, each subtending an azimuthal angle of $45^{\circ}$. To evaluate the physical viability of this model, we consider the epoch of maximum polarization, during which V~CVn optimally illuminates a single $45^{\circ}$ arc. Geometrically, this arc spans $12.5\%$ ($45^{\circ}/360^{\circ}$) of the full azimuth, which means that it intercepts a maximum of $12.5\%$ of incident starlight. Assuming a dust albedo of $0.5$, the cloud reflects $6.25\%$ of the total stellar flux. Even under the assumption that the scattered light achieves a maximum fractional polarization of $30\%$, the net polarized flux is limited to $\sim 1.9\%$ of the total stellar flux. Because V~CVn exhibits an observed polarization amplitude of up to $6\%$, such distant structures lack the geometric leverage and scattering efficiency required to reproduce the signal.

We therefore conclude that a more compact, continuous asymmetric dust shell restricted to a few stellar radii remains the only physically viable mechanism capable of scattering enough light to explain the observed polarimetric variations.
The most robust explanation remains a stable, circumstellar disk or flat envelope, analogous to the structure of $L_2$~Puppis \citep{Kervella2015}. A disk provides a permanent symmetry axis that explains the 60-year stability of the position angle. We further hypothesize that the formation and maintenance of this flattened envelope or disk might be influenced by the high proper motion of V~CVn through the ISM. The resulting ram pressure from this motion could play a significant role in shaping or stabilizing the circumstellar material into the observed geometry.
The variability in the degree of polarization is then explained by the formation and passage of stochastic dust clumps within this disk.

Our observational data and the modeling results reveal a complex phase relationship between the degree of polarization and the optical brightness. While anticorrelation is generally observed (polarization maxima occurring near brightness minima), the phase shifts are notably inconsistent, ranging from a lead or lag of several weeks to near-perfect synchronization in some cycles. We propose that these shifting offsets are a geometric effect of the stochastic emergence of dust clumps relative to the observer's line of sight, as illustrated in Fig. \ref{fig:scetch}. If a newly formed clump is positioned along the line of sight to the stellar photosphere, the brightness starts to darken before the polarization increases. Detailed 3D hydrodynamical simulations of evolved giant star atmospheres and interferometric imaging show strongly evolving and structured atmospheres to motivate our model of stochastic clumps \citep{Hofner2022}\citep{Ohnaka2016}. As the clump moves laterally—either due to orbital motion within the disk or expansion—it eventually reaches a scattering geometry that maximizes the observed polarization. The direction and magnitude of the phase shift thus depend on the specific projected trajectory and velocity of the clump relative to the photosphere. This "clumpy" dynamical model naturally accounts for the erratic timing of the polarization peaks that a purely pulsation-driven, symmetric shell model cannot explain.
Hypothetically, the observed chromatic dependence of the polarization angle may arise from the superposition of multiple such stochastic scattering components. In our model, this would reflect that individual dust clumps with diverse grain properties contribute unevenly across the spectrum. A more detailed investigation of these wavelength-dependent effects—including further model refinement and spectropolarimetric monitoring—is deferred to future work.

An intensification of dust production is likely triggered by the star’s approximately 15-year secondary period (LSP), leading to episodic increases in mass loss. This combination of a stable disk geometry and LSP-driven clumps fully accounts for the observed phenomena.

\section{Conclusions}

In this study, we presented a multi-technique analysis of the semi-regular variable (SRV) V~CVn, combining high-angular-resolution interferometry with long-term photopolarimetry and spectropolarimetry. Our findings lead to the following key conclusions:

\begin{itemize}
     \item For the first time, interferometric observations have resolved V~CVn and its immediate circumstellar environment. These measurements provide direct evidence of the stellar disk and the proximal dusty shell.
    
     \item Long-term polarimetric measurements that span several decades reveal a remarkably stable orientation of the polarization position angle. This indicates the presence of a permanent, geometrically locked asymmetric structure—such as a disk or a flattened envelope—rather than transient atmospheric features.

     \item The observed episodes of high polarization pulsation cycles are consistent with enhanced dust production occurring during specific phases of a proposed Long Secondary Period (LSP). This periodic increase in dust density within the stable asymmetric geometry provides a coherent explanation for the significant variability in the degree of polarization while the position angle remains fixed.
    
    \item Our proposed phenomenological model for the origin of the polarization, which incorporates large clumps injected stochastically into this stable asymmetry, is capable of reproducing the observed phenomena. This accounts for both the high degree of variable polarization anti-correlating with brightness, the constant $PA$ range observed over time, and phase shifts that vary between leads, lags, and synchronization across different cycles, and can be explained by the trajectory of these clumps relative to the observer's line of sight.

\end{itemize}

Although our results provide a first framework for understanding V~CVn, more interferometric observations, imaging, and detailed physical modeling of the variability of the  polarization are still necessary to fully substantiate these hypotheses and decipher the precise dynamics of the mass-loss processes involved.

\section*{Data availability}
Table \ref{table:new_obs} is available in electronic form at the CDS via anonymous ftp to cdsarc.u-strasbg.fr (130.79.128.5) or via http://cdsweb.u-strasbg.fr/cgi-bin/qcat?J/A+A/.

\begin{acknowledgements}
The lead researcher for spectropolarimetry analysis is Yanko Nikolov.
Corresponding author and leading role in photopolarimetry and coordination of interferometric observations: Nicolaus Steenken.
Corresponding author and leading role in theoretical modeling: Richard Ignace.
This work is based on observations obtained with the Georgia State University Center for High Angular Resolution Astronomy Array at Mount Wilson Observatory.  The CHARA array is supported by the National Science Foundation under Grants Nos. AST-2034336 and AST-2407956. Institutional support has been provided from the GSU College of Arts and Sciences and the GSU Office of the Vice President for Research and Economic Development. Some of the CHARA data was collected through the Fast Turnaround Snapshot Imaging Mode.
This research has used the services of the Jean-Marie Mariotti Center (Aspro, SearchCal, OIFits Explorer, LitPro, OImaging and GetStar).
YN acknowledges support by grant: K$\Pi$-06-M88/1 "Interstellar medium - 3D distribution and physical characteristics", by the Bulgarian National Science Fund. 
NS thanks John Simpson (Parton, Scotland) for his polarization measurements and Isabelle Tallon-Bosc (University of Lyon) and Josef Hron (University of Vienna) for their support in analyzing the CHARA data.
SK acknowledges funding for MIRC-X from the European Research Council (ERC) under the European Union's Horizon 2020 research and innovation program (Starting Grant No. 639889 and Consolidated Grant No. 101003096).
JDM acknowledges funding for the development of MIRC-X (NASA-XRP NNX16AD43G, NSF-AST 2009489) and MYSTIC (NSF-ATI 1506540, NSF-AST 1909165).

\end{acknowledgements}

%%%%%%%%%%%%%%%%%%%%%%%%%
\bibliographystyle{aa} % style aa.bst
\bibliography{vcvn_bib_2} % your references Yourfile.bib
%%%%%%%%%%%%%%%%%%%%%%%%%

%Appendix
%%%%%%%%%%%%%%%%%%%%%%%%%
\begin{appendix}
\section{ Introduction}\label{Introduction}
Notes Table 1:

(c) The COBE DIRBE IR light curves \citep{2004yCat..21540673S} were converted to magnitudes using:
\begin{equation}
m_{\text{DIRBE}} = -2.5 \cdot \log_{10}\left(\frac{F_\nu}{F_0}\right)
\end{equation}
where $F_0$ is the zero-point flux corresponding to $0\,\text{mag}$ ($F_{0, \text{H}} = 1068\,\text{Jy}$ and $F_{0, \text{K}} = 660\,\text{Jy}$).

\vspace{1cm}
(f) Bolometric luminosity was calculated using the distance modulus equation:
\begin{equation}
\log(L/L_\odot) = 0.4 \cdot (M_{\odot,\text{bol}} - V + 5 \log d - 5 + A_V - BC_V)
\end{equation}
using:
\begin{itemize}
  \item $V$-magnitude range: $6.5$ to $8.6\,\text{mag}$
  \item Gaia distance $d = 499\,\text{pc}$
  \item Interstellar extinction $A_V \approx 0$ (assumed negligible)
  \item Solar bolometric magnitude $M_{\odot,\text{bol}} = 4.74\,\text{mag}$
  \item Variable bolometric correction $BC_V$ ranging from $-2.2$ to $-2.8\,\text{mag}$, corresponding to the spectral type variation (M4e--M6eIIIa).
\end{itemize}
\section{ CHARA interferometry}\label{CHARA}
%-------------------------------------------------------------

%-------------------------------------------------------------
\begin{table}[!ht]
\caption{Calibrator stars and adopted uniform disk diameters in the H- and K-bands \citep{2017yCat.2346....0B}.} \label{tab2}
    \centering
    \begin{tabular}{ l l l l l }
    \hline
        HD & Type & Distance$^{a}$ (pc) & UDD H$^{b}$ & UDD K$^{b}$ \\ \hline\hline
        120348 & K1III & 185.7 & 0.712 & 0.716 \\ 
        152949 & K0 & 301.7 & 0.341$^{c}$ \\
        153144 & K0 & 274.8 & 0.371$^{c}$ \\
        116172 & K0III & 186.1 & 0.545 & 0.547 \\
        181122 & K1III & 161.8 & 0.763 & 0.716\\\hline
    \end{tabular}            
 \label{calibrators}
 \raggedright
 \small
 Notes: $^{a}$\cite{2023A&A...674A...1G} $^{b}$ (mas) $^{c}$ I-band
\end{table}

\section{Interstellar polarization toward V CVn}\label{ISP_toward_VCVn}
%--------------------------------------------------------------------------------------------------------

Spectropolarimetric observations of V CVn and stars in its direction were obtained between 10 June 2024 and 2 May 2025. A summary of the observations is given in Table \ref{tab.obsjournalspp}, which lists object, date, UT, total exposure time, angular distance from V CVn, distance and, in the last two columns, observed degree of polarization $P_{\rm V,\mathrm{obs}}$ and position angle $PA_{\rm V,\mathrm{obs}}$ in the synthetic V band.

\begin{table*}
\centering
\caption{Journal of spectropolarimetric observations of V CVn and the stars of its direction.}
\begin{tabular}{ l  c  c  c  c c c c c c }
\hline
\hline
Object & Date   & UT & $Exp.^{a}$(s)  &  Angular   & $Distance^{b}$ & $P_{\rm V,\mathrm{obs}}$ & $PA_{\rm V,\mathrm{obs}}$ \\
       &        &    &                &  distance  & pc       &  (\%)	& [deg.]     \\
\hline
V CVn 	&2024-06-10~&~19:56~&~192~&--&~$499^{+25}_{\rm -26}$~&~$1.26 \pm 0.02$~&~$112.0\pm0.5$ \\
		    &2024-07-05~&~20:02~&~288~&  &		             &~$2.24 \pm 0.02$~&~$97.0\pm0.3$	\\
            &2024-12-31~&~02:41~&~256~&  &                   &~$1.48 \pm 0.02$~&~$86.9\pm0.4$	\\
		    &2025-05-02~&~19:47~&~288~&  &		             &~$1.72 \pm 0.01$~&~$86.0\pm0.2$  \\

\hline
  TYC         &~2024-07-05~&~21:27~&~2400~&~8.5'~&~$583^{+4}_{\rm -4}$~&~$0.19 \pm 0.05$~&~$113.3 \pm 7.9$\\
  3460-1345-1 &&&&&&&& \\                    
\hline
 TYC          &~2024-06-10~&~20:26~&~2400~&~5.9'~&~$587^{+4}_{\rm -4}$~&~$0.46 \pm 0.09$~&~$114.5 \pm 4.5$  \\
 3460-2045-1  &&&&&&&& \\ 
\hline
 BD +46 1863  &~2024-06-10~&~21:22~&~2400~&~2.5'~&~$1597^{+27}_{\rm -30}$~&~$0.27 \pm 0.03$~&~$104.0 \pm 3.2$  \\
                 &~2024-07-05~&~20:30~&~1920~&      &                    &~$0.08 \pm 0.04$~&~$76.7 \pm 9.5$   \\

\hline
\hline

\label{tab.obsjournalspp}
\end{tabular} 
\\
Note: $^{a}$ - Total exposure time; $^{b}$ Distance based on Gaia EDR3 \citep{2021AJ....161..147B}.  
\end{table*}

\subsection{Polarization vs. distance}

The variation in the degree of polarization P(\%) with distance d (in kpc) is modeled by the two equations: $P(\%) \approx 0.13+1.81d-0.47d^2+0.036d^3$ \citep{2002ApJ...564..762F} and $P(\%) \approx 0.087+1.526d-0.396d^2+0.033d^3$ \citep{2023AJ....165...87V}. Both equations describe the trend of less than linear gains in polarization amplitude at large distances \citep[see also][]{2025Ap&SS.370...57I}. At the same time, the non-zero constant terms imply an initial degree of polarization of 0.13\% and 0.087\%, respectively, at a distance of 0~pc. 
Because the cubic polynomial is not based on physical arguments, its coefficients do not have a clear physical meaning \citep{2023AJ....165...87V}.

%-------------------------------------------------------------
   \begin{figure}
   \centering
   \includegraphics[width=\linewidth]{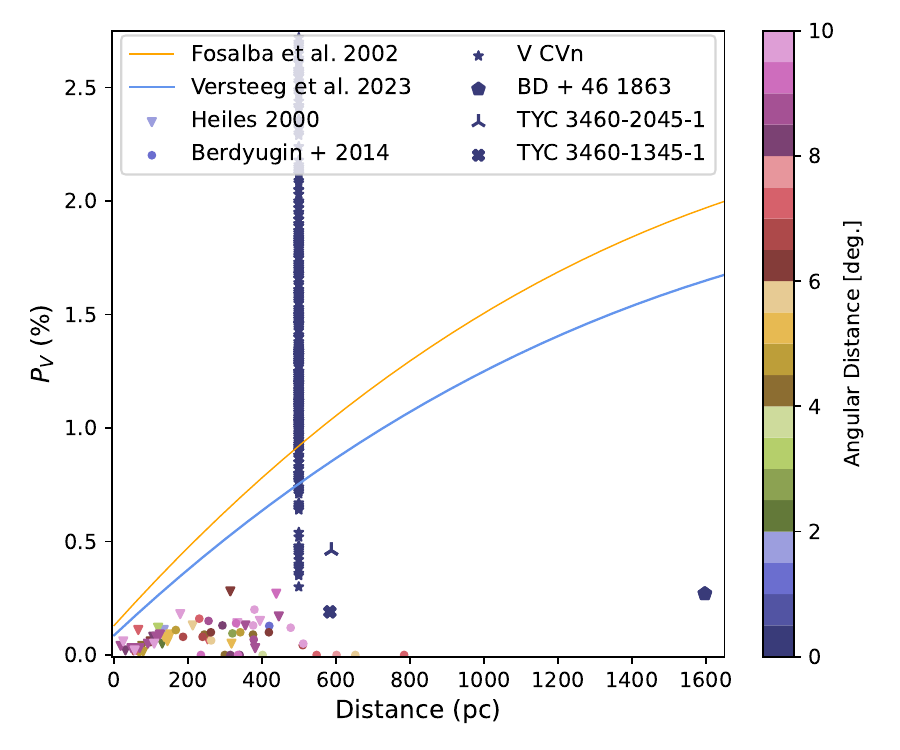}
   \vspace*{0cm}
      \caption{Polarization vs. distance. The orange and the blue lines represent the polarization vs. distance relationship taken from \citep{2002ApJ...564..762F} and \citep{2023AJ....165...87V}, respectively. The color of each star indicates its angular distance from V CVn.}
         \label{distance_pol}
   \end{figure}
%----------------------------------------------------------------

In Fig. \ref{distance_pol}, we show stars within a $10^\circ$ angular distance from V CVn. Data taken from \citet{2000AJ....119..923H} (represented by triangles) and \citet{2014A&A...561A..24B} (represented by circles). The stars, listed in Table \ref{tab.obsjournalspp} are presented with different symbols. Most stars in both catalogs are located within a distance of approximately 500 pc, based on Gaia EDR3 measurements \citep{2021AJ....161..147B}. A general trend of increasing polarization  is observed with distance. 

In particular, the degree of polarization of BD +46 1863 in the synthetic V band is comparable to that of stars located within 500 pc, suggesting that its polarization is likely caused by foreground dust at distances up to $\approx$ 500 pc. The observed stars, listed in Table~\ref{tab.obsjournalspp}, are located at distances ranging from 499 to 1597 pc and lie within an angular distance from V CVn of approximately 9 arcminutes.

Historically, V CVn has exhibited strong variability in its polarization. The lowest recorded value in the V-band is $P_{\rm obs}$ = 0.30\% at a position angle of $PA_{\rm obs}$ = $98.8^\circ$ (observed on 7 August 2020) \cite{Neilson2023}, while the highest value is $P_{\rm obs}$ = 6.94\% at $PA_{\rm obs}$ = $100.5^\circ$ by the HPOL collaboration \citep{2012AIPC.1429..226M, Davidson2014}.

\begin{figure*}[t]
    \centering
    \includegraphics[width=\textwidth]{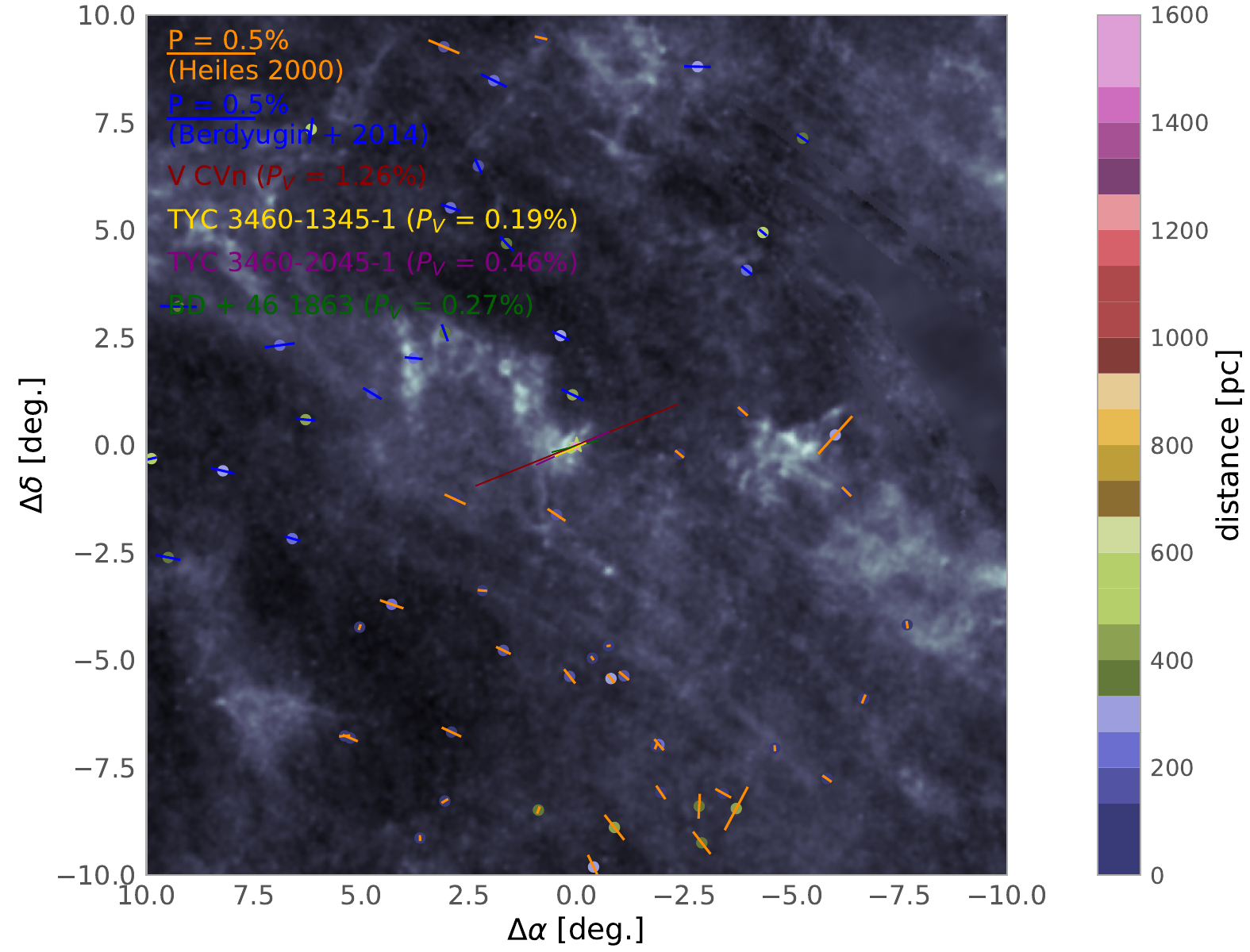} 
    \caption{The interstellar polarization of the field stars around  V CVn, with orange for \citet{2000AJ....119..923H}, and blue for \citet{2014A&A...561A..24B}. The degree of polarization is proportional to the length of its bar. The horizontal bar of the top right presents 0.5\% polarization. The color of every star corresponds to its distance. The background image represents 100~$\mu$m dust emission maps \citep{1998ApJ...500..525S}.} 
    \label{3D_Pol}
\end{figure*}

A 3D map of interstellar polarization in the field around V CVn is presented in Fig. \ref{3D_Pol}. Data are taken from \citet{2000AJ....119..923H,2014A&A...561A..24B} and this work. The position angle of V CVn is similar to the position angles of the stars presented in Table \ref{tab.obsjournalspp}. The color of every star corresponds to its distance. The background image represents 100 $\mu$m dust emission maps \citep{1998ApJ...500..525S}. 
Figures \ref{distance_pol} and \ref{3D_Pol} provide information on the 3D distribution of the degree of polarization and position angle in the field around V CVn.

\subsection{Determining the Serkowski law parameters describing the interstellar polarization toward V CVn.}

The Serkowski's law describes the wavelength dependence of the degree of linear interstellar polarization \citep{1975ApJ...196..261S}:
\begin{equation}
P_{\rm isp}(\lambda ) = P_{\rm max}\exp \left(-K\ln ^{2}\frac{\lambda _{\rm max}}{\lambda }\right),
\end{equation}
where $P_{\rm max}$ is the peak of the interstellar polarization at wavelength $\lambda_{\rm max}$.

Two methods are commonly used to determine the interstellar polarization of the object of interest: the field stars method \citep{2023A&A...679A.150N,2022NewA...9701859N} and a method based on depolarization effects in the emission lines  \citep[e.g.,][]{1990ApJ...365L..19S, 2025A&A...697A.237S}. 
The field stars method requires stars, at similar distance and small angular distance around the V CVn.

At low polarization degrees, fitting the degree of polarization with the Serkowski law and determining its parameters becomes challenging. This difficulty increases for fainter stars and noisier data. However, if the position angle does not vary with wavelength, satisfactory accuracy of the position angle can be achieved using the SBAND procedure in IRAF and the derivation of the position angle through a synthetic V filter.

In cases of low interstellar polarization, we employ the following methodology to determine the parameters of Serkowski’s law. We consider two stars (TYC 3460-1345-1 and BD +46 1863) with similar observed degrees of polarization and assume that their interstellar polarization follows the empirical Serkowski law. We used a model that reproduces the Serkowski profile and computed the minimum of the sum of absolute differences between the model and the observed polarization values. We compute in the range from $\lambda=5000~\AA$ to $\lambda=8200~\AA$ (we exclude data with $\lambda < 5000~\AA$ due to high noise levels in this wavelength range). Spectropolarimetric observations of TYC 3460-2045-1 were excluded from the analysis due to excessive noise.

The parameter space is explored using a three-dimensional grid over the Serkowski parameters: maximum polarization $P_{\rm max}$, the wavelength of maximum polarization $\lambda_{\rm max}$, and the width parameter $K$. The search is constrained to the following ranges: $0.10 < P_{\rm max}(\%) < 0.55$, while $\lambda_{\rm max}$ and $K$ vary within their respective 1$\sigma$ uncertainty intervals (the mean of $\lambda_{\rm max}$ is $\lambda_{\rm max}=0.58$ $\mu$m with a standard deviation 
$\sigma_{\rm \lambda_{\rm max}}=0.07$ $\mu$m and the mean of the K coefficient is $K=1.23$ with $\sigma_{\rm K}=0.32$ \citep{Nikolov2026}. The degree of polarization is debiased before fitting. The best-fit parameters are those that minimize the value of the selected fitting criterion. 

\begin{figure}
   \centering
   \includegraphics[width=\linewidth]{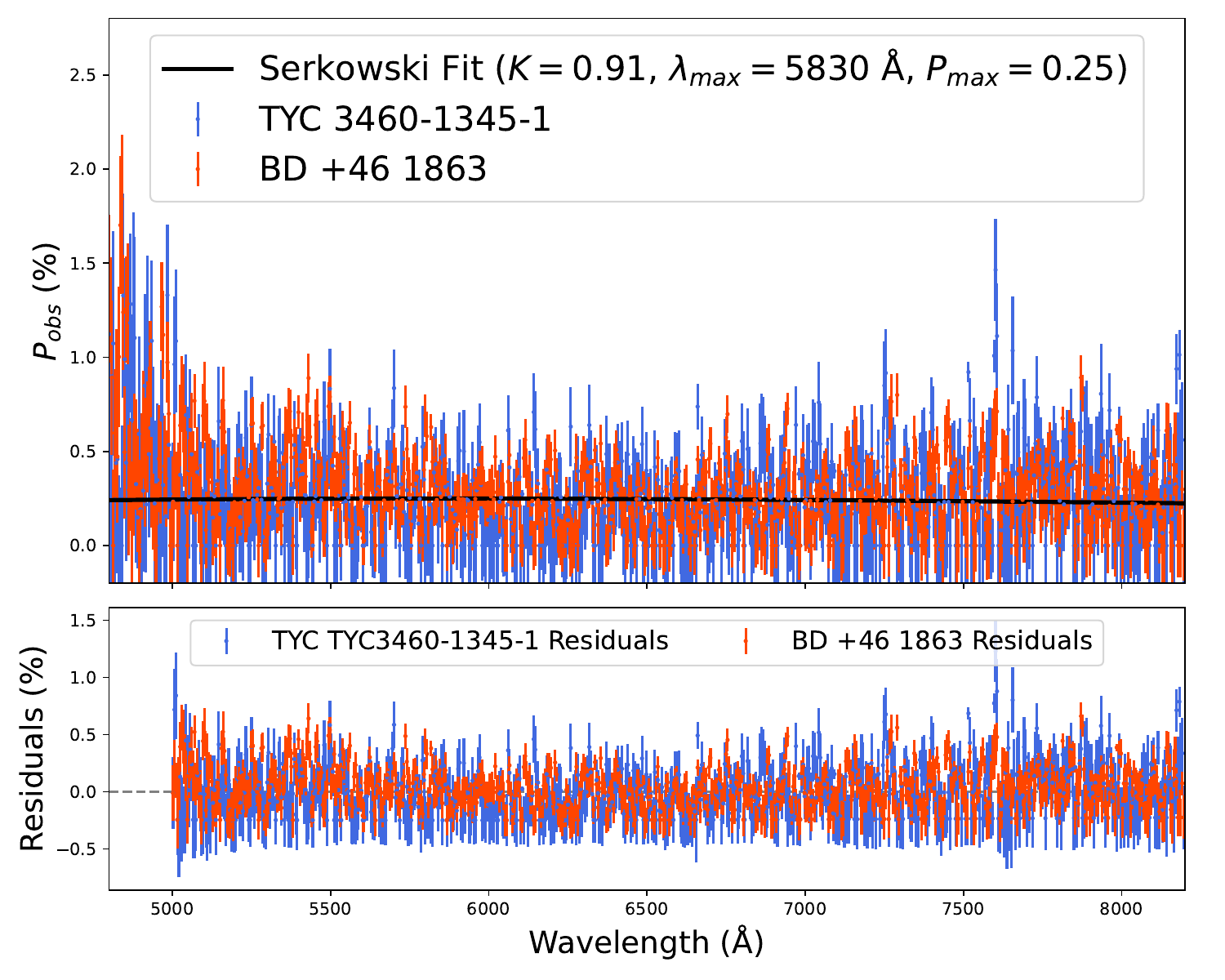}
   \vspace*{0cm}
      \caption{Degree of polarization of TYC 3460-1345-1 and BD +46 1863, best fit model (top) and residuals.}
         \label{bst.fit}
   \end{figure}

Based on this methodology, we calculate $P_{\rm max}(isp)=0.25 \%$; $K=0.91$ and $\lambda{max}=5830~\AA$. The $PA$ of the interstellar polarization is $PA_{\rm isp}=110.6~deg.$ This position angle represents the average values of the position angles of TYC 3460-1345-1; TYC 3460-2045-1, and BD +46 1863 (the position angle obtained on 2024-06-10) in synthetic V band. In Fig. \ref{bst.fit} the degree of polarization of TYC 3460-1345-1 and BD +46 1863, best fit model (top), and residuals (bottom) is shown. The mean Stokes parameters were determined with uncertainties corrected for by Student (t), and the interstellar polarization and its confidence range were derived from the corresponding confidence ellipse following \cite{Clarke2010}.

We calculated the Stokes parameter of the interstellar polarization $Q_{\rm isp}(\lambda)$ and $U_{\rm isp}(\lambda)$ using the following equations: 
\begin{eqnarray}
Q_{\rm isp}(\lambda)=P_{\rm isp}(\lambda)cos(2\times PA_{\rm isp}),\\
U_{\rm isp}(\lambda)=P_{\rm isp}(\lambda)sin(2 \times PA_{\rm isp}).
\end{eqnarray}

Table~\ref{fit.coefficients} lists the coefficients obtained from the polynomial fits to the intrinsic position angle and from the power-law fits to the intrinsic polarization. These coefficients quantify the wavelength dependence of $PA_{\rm intr}$ and $P_{\rm intr}$ for V~CVn.

\begin{table}[t]
\centering
\caption{Fitting coefficients for $PA_{\rm intr}$ and $P_{\rm intr}$ (\%).}
\label{fit.coefficients}
\begin{tabular}{c|c|c|c|c|c}
\hline
Date-obs. & \multicolumn{3}{c|}{{$PA_{\rm intr}$ fitting coeff.}} & \multicolumn{2}{c|}{{$P_{\rm intr}$ fitting coeff.}} \\
 & $a$ & $b$ & $c$ & $p_0$ & $\beta$ \\
\hline
2024-06-10 & 148.3 & -53.5 & 13.5  & 1.04 & -0.87 \\
\hline
2024-07-05 & 47.4  & 86.9  & -37.4 & 1.96 & -1.57 \\
\hline
2024-12-31 & 34.7  & 91.9  & -39.3 & 1.40 & -1.88 \\
\hline
2025-05-02 & 121.6 & -79.6 & 41.2  & 1.51 & -1.71 \\
\hline
\end{tabular}
\end{table}

\section{ Photopolarimetry}\label{Photopol}

\subsection{Calibration and Error Calculation}

The errors of intrinsic Stokes $Q_{\rm intr}$ and $U_{\rm intr}$  in Table~\ref{tab3} were calculated according to:

\begin{equation}
\sigma Q_{\rm intr}={\sqrt{\sigma_{\rm Q_{\rm cal}}^2+\sigma_{\rm Q_{\rm isp}}^2}}
\end{equation}

\begin{equation}
\sigma U_{\rm intr}={\sqrt{\sigma_{\rm U_{\rm cal}}^2+\sigma_{\rm U_{\rm isp}}^2}}
\end{equation}

The errors of $P_{\rm intr}$ and $PA_{\rm intr}$ were calculated as follows:

\begin{equation}
\sigma P_{\rm intr} = \frac{\sqrt{Q_{\rm intr}^2 \sigma_{Q_{\rm intr}}^2 + U_{\rm intr}^2 \sigma_{U_{\rm intr}}^2}}{P_{\rm intr}}
\end{equation}

\begin{equation}
\sigma {PA_{\rm intr}} = \frac{\sqrt{Q_{\rm intr}^2 \sigma_{U_{\rm intr}}^2 + U_{\rm intr}^2 \sigma_{Q_{\rm intr}}^2}}{2P_{\rm intr}^2}
\end{equation}

The standard errors of the mean values of the observed and calibrated Stokes Q, U were calculated according to \cite{Clarke2010} using the student-t distribution for 1-sigma confidence limits.
\begin{figure*}
    \centering
    \includegraphics[width=\textwidth]{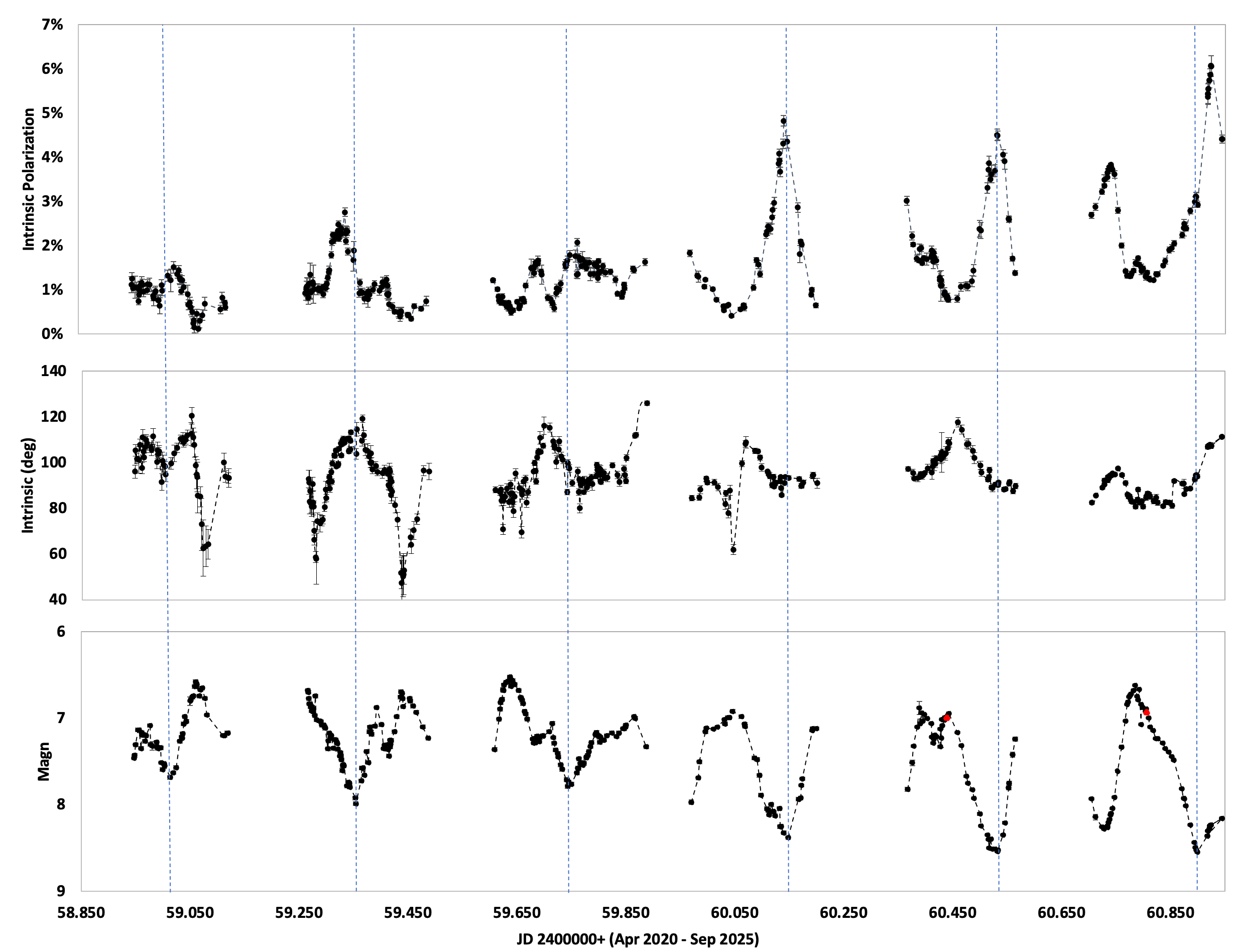} 
    \caption{Six pulsation cycles of the intrinsic polarization of V CVn (upper panel). The intrinsic polarization angel (PA) is shown in the middle panel and the brightness in the lower panel. All observations made in V-band. Brightness minima are marked with dotted lines. Brightness and polarization are anticorrelated, but there are time shifts between the brightness minima and the polarization maxima like in the second cycle from left (2021) the third (2022) and the last pulsation cycle (2025). Red points in the light curve mark the epochs of our CHARA array observations.}
    \label{IntrinsicPol}
\end{figure*}

\begin{figure}
   \centering
   \includegraphics[width=\linewidth]{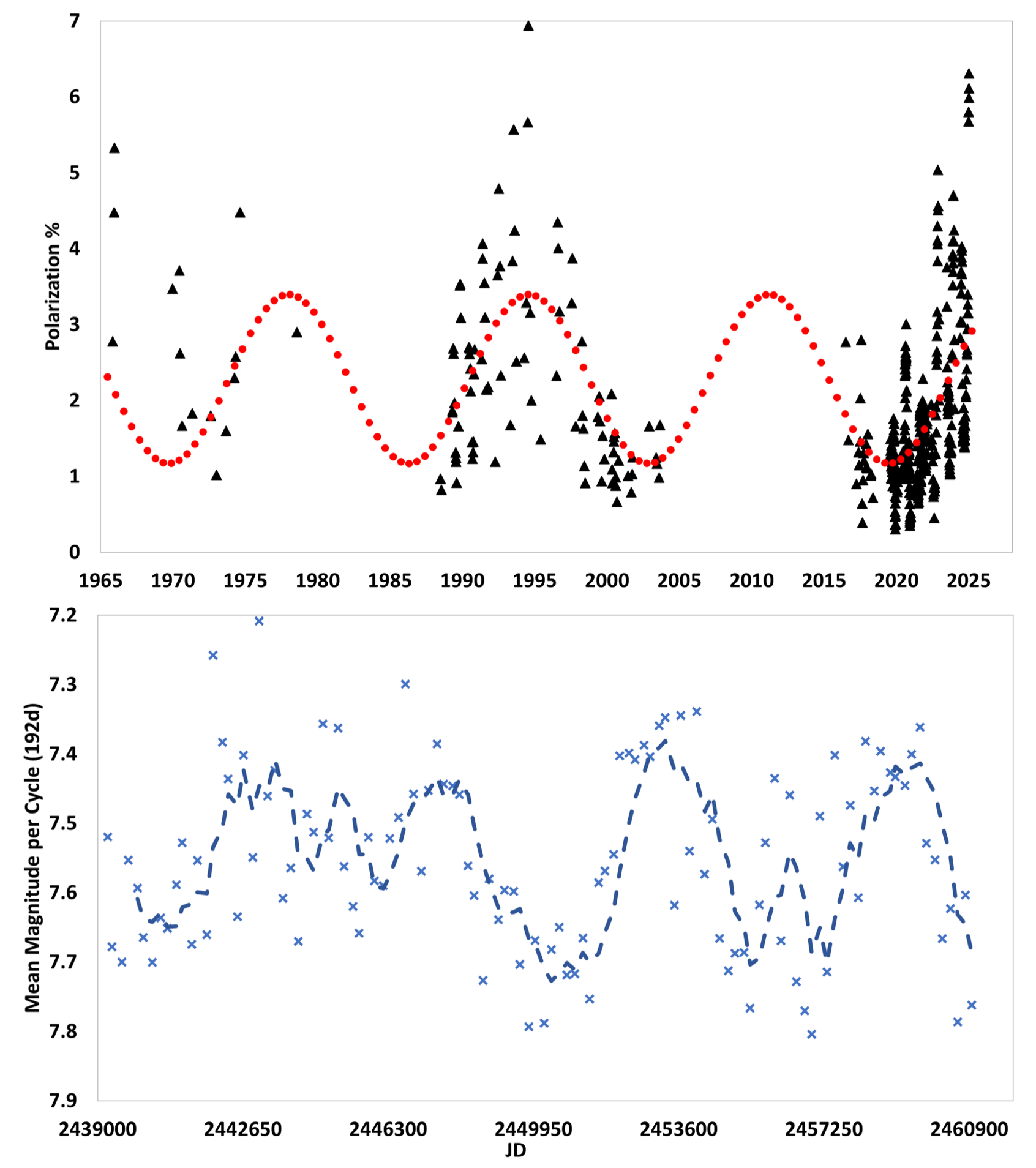}
     \caption{Comparison between $V$-band linear polarization percentage (top panel) and visual light curve (bottom panel) over six decades. Top: Black triangles represent linear polarization measurements compiled from \cite{Serkowski2001}, \cite{Davidson2014}, \cite{Safonov2019} and this work. The red dotted curve shows a sinusoidal model of maximum polarization variation with a 15-year period (amplitude ranging between 1\% and 5\%).
     Bottom: Mean visual magnitudes per pulsation cycle ($192\text{ d}$, blue crosses) derived from AAVSO observations using VStar. The blue dashed line indicates a 5-cycle moving average. Time-dependent period and amplitude variations were extracted using the Weighted Wavelet Z-transform (WWZ) technique.
     } 
    \label{Historic}
   \end{figure}
High polarization standard stars such as $\phi$~Cas and 2H~Cam were measured in agreement with the published values of \cite{Hsu1982} as shown in Table~\ref{tab2}.
\begin{table}[!htbp]
\caption{Highly polarized stars measurements and comparison to reference} \label{tab2}
    \centering
    \begin{tabular}{ l l l l l }
    \hline
        Star & JD & $P_{\rm cal}$  & $PA_{\rm cal}$ & Observer  \\ \hline\hline
        $\phi$~Cas & 2460538.4 & 3.33\% & 92.5 & 1 \\ 
        $\phi$~Cas & 2460601.3 & 3.18\% & 90.1 & 1 \\ 
        $\phi$~Cas & 2460605.3 & 3.35\% & 91.8 & 1 \\ 
        $\phi$~Cas & 2460677.3 & 3.35\% & 91.1 & 1 \\ 
        $\phi$~Cas & 2460712.3 & 3.41\% & 92.3 & 1 \\ 
        $\phi$~Cas & 2460888.4 & 3.35\% & 92.2 & 1 \\ 
        $\phi$~Cas & 2460918.3 & 3.37\% & 91.8 & 1 \\ 
        $\phi$~Cas & 2460963.3 & 3.30\% & 91.0 & 1 \\ \hline\hline
        $\phi$~Cas & Average  & 3.33\% & 91.6 & 1 \\ \hline\hline
        ~ & ~ & ~ & ~ & ~ \\ \hline
        $\phi$~Cas & 2460729.5 & 3.36\% & 92.8 & 2 \\ 
        $\phi$~Cas & 2460734.3 & 3.25\% & 92.4 & 2 \\ 
        $\phi$~Cas & 2460743.4 & 3.40\% & 93.4 & 2 \\ 
        $\phi$~Cas & 2460747.4 & 3.37\% & 93.0 & 2 \\ 
        $\phi$~Cas & 2460753.4 & 3.27\% & 92.9 & 2 \\ 
        $\phi$~Cas & 2460755.4 & 3.27\% & 92.6 & 2 \\ 
        $\phi$~Cas & 2460760.4 & 3.23\% & 93.4 & 2 \\ 
        $\phi$~Cas & 2460767.4 & 3.21\% & 92.6 & 2 \\ 
        $\phi$~Cas & 2460789.5 & 3.27\% & 92.7 & 2 \\ 
        $\phi$~Cas & 2460790.5 & 3.26\% & 92.3 & 2 \\ 
        $\phi$~Cas & 2460796.5 & 3.29\% & 92.5 & 2 \\ 
        $\phi$~Cas & 2460804.5 & 3.30\% & 92.6 & 2 \\ 
        $\phi$~Cas & 2460812.5 & 3.30\% & 92.5 & 2 \\ 
        $\phi$~Cas & 2460818.5 & 3.36\% & 92.7 & 2 \\ 
        $\phi$~Cas & 2460944.4 & 3.26\% & 92.7 & 2 \\ \hline
        $\phi$~Cas & Average  & 3.29\% & 92.8 & 2 \\ \hline
        $\phi$~Cas & \cite{Hsu1982} & 3.34\% & 92.3 & ~ \\ \hline
        ~ & ~ & ~ & ~ & ~ \\ \hline
        2H Cam & 2460411.3 & 3.45\% & 116.6 & 1 \\ 
        2H Cam & 2460775.3 & 3.50\% & 114.6 & 1 \\ \hline
        2H Cam & Average   & 3.47\% & 115.9 & 1 \\ \hline
        ~ & ~ & ~ & ~ & ~ \\ \hline
        2H Cam & 2460553.5 & 3.40\% & 116.7 & 2 \\ 
        2H Cam & 2460804.4 & 3.44\% & 115.0 & 2 \\ \hline
        2H Cam & Average   & 3.42\% & 116.5 & 2 \\ \hline
        2H Cam & \cite{Hsu1982} & 3.49\% & 116.6 \\ \hline
    \end{tabular}
\end{table}

\begin{table}
\caption{Unpolarized stars used as standards, and their polarization according to \cite{Schmidt1992}\label{tab3}}
\centering
\begin{tabular}{l  c  c  c  c  c  c} 
 \hline
 Star & $mag(V)$ &$P$ (\%) & $\sigma_P$ (\%) & Observer \\ \hline \hline
 $\beta$~Cas & 2.3 & 0.037 & 0.024& 1,2\\
 $\theta$~UMa & 3.2 & 0.037 & 0.014& 1,2\\ 
 $\beta$~UMa & 2.4& 0.009 & 0.019 &2\\
 HD 21447 & 5.1 & 0.051 & 0.020& 1\\ 
 \hline
\end{tabular}
\end{table}

\begin{table*}
\caption{Polarimetric and photometric observations obtained between 2023 and 2025.}
\label{table:new_obs}
\centering
\begin{tabular}{ccccccccccc}
\hline\hline
JD & $P_{\text{cal}}$ & $PA_{\text{cal}}$ & $Q_{\text{cal}}$ & $\sigma Q_{\text{cal}}$ & $U_{\text{cal}}$ & $\sigma U_{\text{cal}}$ & mag(V) & $\sigma_{\text{mag}}$ & Ob \\
($-2\,400\,000$) &  & (deg) &  &  &  &  & (mag) & (mag) & \\
\hline
59970.4 & 0.0199 & 87.3 & -0.0198 & 0.0004 & 0.0019 & 0.0008 & 7.97 & 0.03 & 2 \\
59983.4 & 0.0150 & 88.3 & -0.0149 & 0.0005 & 0.0009 & 0.0005 & 7.69 & 0.02 & 1 \\
59985.4 & 0.0149 & 91.5 & -0.0149 & 0.0012 & -0.0008 & 0.0012 & 7.50 & 0.02 & 1 \\
59996.4 & 0.0129 & 96.1 & -0.0126 & 0.0005 & -0.0027 & 0.0005 & 7.16 & 0.02 & 1 \\
59998.5 & 0.0144 & 94.3 & -0.0142 & 0.0002 & -0.0022 & 0.0005 & 7.11 & 0.01 & 2 \\
60011.4 & 0.0123 & 94.9 & -0.0121 & 0.0003 & -0.0021 & 0.0006 & 7.12 & 0.01 & 2 \\
60018.4 & 0.0098 & 94.1 & -0.0097 & 0.0003 & -0.0014 & 0.0006 & 7.10 & 0.01 & 2 \\
60031.4 & 0.0079 & 89.5 & -0.0079 & 0.0003 & 0.0001 & 0.0007 & 7.01 & 0.01 & 2 \\
60032.3 & 0.0073 & 93.8 & -0.0073 & 0.0004 & -0.0010 & 0.0004 & 7.07 & 0.02 & 1 \\
60037.4 & 0.0078 & 86.3 & -0.0077 & 0.0003 & 0.0010 & 0.0006 & 6.99 & 0.01 & 2 \\
\hline
\end{tabular}
\tablefoot{Only the first ten rows are shown for guidance. The complete merged catalog, including the previous measurements from \citet{Neilson2023} (CDS J/A+A/677/A96), is available in electronic form at the CDS via anonymous ftp to cdsarc.u-strasbg.fr (130.79.128.5) or via http://cdsweb.u-strasbg.fr/cgi-bin/qcat?J/A+A/..}
\end{table*}

\subsection{Extra long secondary period}
\begin{figure}
   \centering
   \includegraphics[width=\linewidth]{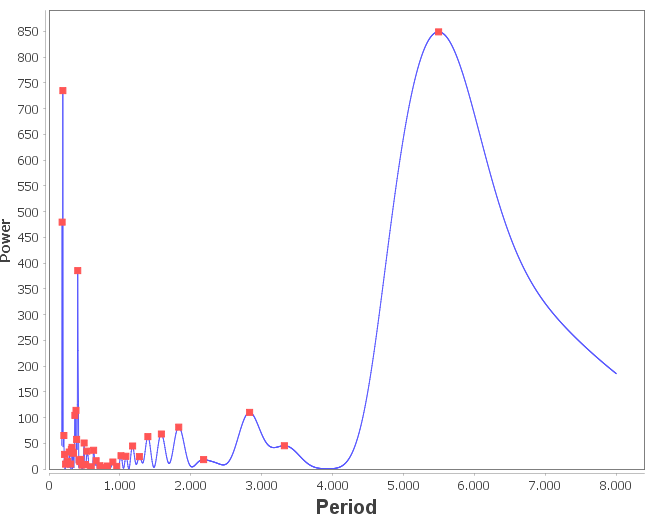}
   \vspace*{0cm}
      \caption{Periodogram of AAVSO visual observations since 1966, derived using the VStar DC DFT algorithm. The primary period is detected at $P = 5495\text{ d}$ ($15.05\text{ yr}$, spectral power 849). Secondary peaks are observed at $P = 195\text{ d}$ (power 734) and $P = 185\text{ d}$ (power 479).}
         \label{Period}
   \end{figure}

A comparison between the long-term $V$-band polarization curve since 1965 and the visual light curve reveals a distinct multi-year modulation. The periodogram analysis of AAVSO visual observations (Fig. \ref{Period}) clearly isolates an extra-long secondary period of $\sim 15\text{ yr}$ ($P = 5495\text{ d}$), a long-term feature previously noted in the literature \citep{Cadmus2024, Poliakova1989}. When comparing this $15$-year photometric baseline with the polarization data, a general anti-correlation is observed, where minimum visual brightness roughly corresponds to maximum polarization degrees as shown in Fig. \ref{Historic}. However, testing this relationship over six decades is hindered by substantial observational coverage gaps in the polarimetric timeline—most notably around 1980 and 2010— during expected polarization maxima. Continuous, high-cadence polarimetric monitoring over future cycles will therefore be essential to firmly confirm this long-term anti-correlation.

\end{appendix}

\end{document}